\documentclass[twocolumn]{aastex631}

\usepackage{multirow}

\usepackage{amsmath}

\usepackage{bm}

\usepackage[normalem]{ulem}

\definecolor{GF}{rgb}{0.5,0,1}
\definecolor{Hiro}{rgb}{0,0,1}
\definecolor{SSH}{rgb}{1,0,0.5}

\begin{document}

\title{Probing the Diversity of Type Ia Supernova Remnants in 3-D Hydrodynamic Simulations with X-ray Spectral Synthesis}

\author[0009-0007-8699-8973]{Yusei Fujimaru}
\newcommand{\KyotoU}{Department of Astronomy, Kyoto University, Kitashirakawa, Oiwake-cho, Sakyo-ku, Kyoto 606-8502, Japan}
\affiliation{\KyotoU}

\author[0000-0002-2899-4241]{Shiu-Hang Lee}
\affiliation{\KyotoU}
\affiliation{Kavli Institute for the Physics and Mathematics of the Universe (WPI), The University of Tokyo, Kashiwa 277-8583, Japan}

\author[0000-0002-4231-8717]{Gilles Ferrand}
\affiliation{{The University of Manitoba, Department of Physics and Astronomy, Winnipeg, Manitoba, R3T~2N2, Canada}}
\affiliation{RIKEN Center for Interdisciplinary Theoretical and Mathematical Sciences (iTHEMS), Wak\={o}, Saitama, 351-0198 Japan}

\author[0000-0002-7507-8115]{Daniel Patnaude}
\affil{Smithsonian Astrophysical Observatory, 60 Garden Street, Cambridge, MA 02138 USA}

\author[0000-0002-7025-284X]{Shigehiro Nagataki}
\affiliation{Astrophysical Big Bang Laboratory (ABBL), RIKEN Pioneering Research Institute (PRI), Wak\={o}, Saitama, 351-0198 Japan}
\affiliation{Astrophysical Big Bang Group (ABBG), Okinawa Institute of Science and Technology Graduate University (OIST), 1919-1 Tancha, Onna-son, Kunigami-gun, Okinawa 904-0495, Japan}
\affiliation{RIKEN-Berkeley Center, RIKEN iTHEMS, University of California, Berkeley, Berkeley, CA 94720, USA}
\affiliation{RIKEN Center for Interdisciplinary Theoretical and Mathematical Sciences (iTHEMS), Wak\={o}, Saitama, 351-0198 Japan}

\author[0000-0003-3308-2420]{R\"udiger Pakmor}
\affiliation{Max-Planck-Institut f\"{u}r Astrophysik, Karl-Schwarzschild-Str. 1, D-85748, Garching, Germany}

\author[0000-0001-6189-7665]{Samar Safi-Harb}
\affiliation{{The University of Manitoba, Department of Physics and Astronomy, Winnipeg, Manitoba, R3T~2N2, Canada}}

\author[0000-0002-4460-0097]{Friedrich K.\ R{\"o}pke}
\affiliation{Heidelberger Institut f{\"u}r Theoretische Studien, Schloss-Wolfsbrunnenweg 35, 69118 Heidelberg, Germany}
\affiliation{Zentrum f{\"u}r Astronomie der Universit{\"a}t Heidelberg, Institut f{\"u}r Theoretische Astrophysik, Philosophenweg 12, 60120 Heidelberg, Germany}
\affiliation{Zentrum f{\"u}r Astronomie der Universit{\"a}t Heidelberg, Astronomisches Rechen-Institut, M{\"o}nchhofstr.\ 12--14, 69120 Heidelberg, Germany}

\author[0000-0002-1796-758X]{Anne Decourchelle}
\affiliation{Université Paris-Saclay, Université Paris Cité, CEA, CNRS, AIM, 91191, Gif-sur-Yvette, France}

\author[0000-0002-5044-2988]{Ivo R. Seitenzahl}
\affiliation{Research School of Astronomy and Astrophysics, Australian National University, Canberra, ACT 2611, Australia}
\affiliation{Mathematical Sciences Institute, Australian National University, Canberra, ACT 0200, Australia}

\begin{abstract}
Type Ia supernovae (SNe), thermonuclear explosions of white dwarfs in binary systems, are widely used as standard candles owing to the empirical width-luminosity relation of their light curves.
Recent theoretical and observational studies indicate a diversity of progenitor systems and explosion mechanisms.
In the supernova remnant (SNR) phase, the diversity in Fe-K$\alpha$ centroid energies and line luminosities suggests variations in the underlying explosion mechanisms.
X-ray spectra of SNRs, which trace shocked ejecta and the surrounding medium, are crucial diagnostics of progenitor systems and explosion physics. 
Thanks to recent advances in spectroscopy with XRISM, high-resolution X-ray spectroscopy enables 3-D diagnostics, including line-of-sight velocities.
In this study, we perform 3-D hydrodynamic simulations of SNRs from six Type Ia explosion models: two each of pure deflagration, delayed detonation, and double detonation. 
Each model is evolved for 1000 years in a uniform medium, consistently accounting for non-equilibrium ionization.
Our efficient numerical scheme enables systematic parameter surveys in full 3-D.
From these models, we synthesize X-ray spectra with $\sim$1 eV resolution, exceeding XRISM/Resolve's spectral resolution.
This work presents the first calculation of X-ray spectra for Type Ia SNRs derived from 3-D hydrodynamic simulations that follow the evolution self-consistently from the SN phase into the SNR phase.
Our results show inter-model diversity in the X-ray spectra. 
Asymmetric, red- and blueshifted line profiles arise from the 3-D ejecta distributions. 
These findings demonstrate that 3-D SNR modeling can reproduce the observed diversity of Type Ia SNRs and provide qualitative constraints on progenitor systems and explosion mechanisms.
\end{abstract}

\keywords{Supernova remnants --- Supernovae --- methods: numerical}

\section{Introduction} \label{sec:intro}

Supernovae (SNe) are among the most extreme phenomena in the Universe.
Their remnants -- supernova remnants (SNRs) -- arise from interaction between the ejecta and the ambient medium; the resulting shocks heat the plasma and produce thermal X-ray lines that encode elemental abundances and ionization states.

In particular, Type Ia SNe play a crucial role as standard candles for measuring cosmic distances and probing cosmic expansion \citep{riess1998observational,perlmutter1999measurements}, and they are also important as one of the major sources of chemical elements in the Universe.
They are understood to be thermonuclear explosions of white dwarfs (WDs) in binary systems.
However, the progenitor channels (progenitor system scenarios and explosion mechanisms) that lead to these explosions remain uncertain, despite numerous possibilities having been proposed \citep[for a recent review, see][]{ruiter2025type}.
Two main progenitor system scenarios have been suggested: the single-degenerate (SD) scenario, in which a WD forms a binary with a normal star \citep{whelan1973binaries, nomoto1982accreting}, and the double-degenerate (DD) scenario, in which the binary consists of two WDs \citep{Iben1984Supernovaemass, webbink1984double}.
With regard to explosion mechanisms, two possibilities are considered: near-Chandrasekhar mass (near-$M_\text{Ch}$) explosions in which a WD explodes at a mass close to the Chandrasekhar limit, and sub-Chandrasekhar mass (sub-$M_\text{Ch}$) explosions in which the exploding WD has a mass below the Chandrasekhar mass. 
These cases involve different explosion physics and yield distinct nucleosynthetic signatures.
The double detonation model \citep{livne1990successive, fink2010double, guillochon2010surface, pakmor2013helium} is considered one of the most promising sub-$M_\text{Ch}$ explosion mechanisms.
In this model, a helium detonation in the outer layer is triggered by mass accretion from the companion.
The shock waves following the He detonation converge inside the WD core, igniting a secondary carbon detonation and leading to the WD explosion.
Recently, \citet{das2025calcium} reported the detection of a double Ca shell together with a single S shell in SNR 0509--67.5, providing the first direct observational evidence for a double detonation origin.

The remnant left behind after the explosion is a SNR, and even during the SNR phase, considerable diversity is observed among Type Ia SNe.
One example is the distribution of the centroid energy and line luminosity of the Fe-K$\alpha$ emission line.
It has been suggested that the Fe-K$\alpha$ line provides a useful diagnostic for discriminating whether an SNR originates from a Type Ia or a core-collapse (CC) supernova \citep{yamaguchi2014discriminating}.
For Type Ia SNe, the density of iron is relatively low, so ionization proceeds less efficiently, and the centroid energy is expected to fall in a relatively low-energy range.
In \citet{Patnaude2015models}, CC SNR models evolved in stellar wind environment were shown to reproduce observed Fe-K$\alpha$ centroid energies and line luminosities, supporting the view that Fe‑K$\alpha$ properties can discriminate between Type Ia and CC origins.
Observations of Type Ia SNRs show that the centroid energy is spread over 6400--6550\,eV.
The line luminosities also span over three orders of magnitude, further highlighting the diversity.

In SNRs, forward and reverse shocks are formed, heating the ambient medium and the ejecta material, respectively.
The heated plasma emits thermal X-rays, from which one can infer various properties such as elemental abundances and plasma conditions.
In particular, emission lines from ejecta heated by the reverse shock dominate the spectrum of young SNRs, and carry information about the progenitor composition.

A notable feature of SNRs is that they are spatially extended objects, with typical sizes of a few to several tens of parsecs, enabling spatially resolved observations.
Although the interiors of stars and supernova explosions cannot be directly observed, SNRs encode their history, allowing indirect constraints on stellar structures and explosion mechanisms.
Moreover, since SNRs are shaped by the interaction between the ejecta and the surrounding medium, they provide valuable clues for constraining SD vs. DD progenitor scenarios as well as near-$M_\text{Ch}$ vs. sub-$M_\text{Ch}$ explosion mechanisms.

Previous theoretical studies have attempted to model Type Ia SNRs using different approaches.
1-D hydrodynamical simulations combined with X-ray spectral synthesis have been widely applied \citep[e.g.,][]{badenes2003thermal, badenes2005thermal, badenes2006constraints}, providing valuable insight into the ionization states and elemental abundances of shocked ejecta.
However, such 1-D studies inherently neglect spatial and velocity asymmetries, which are expected from multi-dimensional explosion models.
3-D hydrodynamic simulations of SNRs have been developed \citep{ferrand20103d} including non-equilibrium ionization (NEI) conditions and X-ray spectra production \citep{ferrand2012three}, but they have focused on studying the back-reaction of shock-accelerated particles and have not been based on 3-D SN explosion models.
More recently, 3-D hydrodynamical simulations from the SN to the SNR evolution have been performed \citep[e.g.,][]{ferrand2019supernova, ferrand2021supernova, ferrand2022double, ferrand2025role}, and the global morphology of SNRs has been studied in detail in 3-D. However, these typically focused on plasma dynamics and morphology, without calculating observable X-ray spectra.
Thus, a self-consistent 3-D framework that evolves explosion models into SNRs and simultaneously predicts their X-ray emission has been lacking.

Recent advances in observational technology have dramatically improved the energy resolution of X-ray spectroscopy.
The first among these is the XRISM Resolve instrument \citep{ishisaki2022status}, which achieves an energy resolution of about 5\,eV.
Such capabilities allow for highly precise X-ray spectra, enabling detailed studies of plasma states and even 3-D structures through Doppler line shift measurements.

As the next step beyond morphological studies, we perform 3-D simulations of various explosion models in a uniform-density environment, consistently following their evolution into the SNR phase including their gas properties.
We further generate synthetic X-ray spectra with high spectral resolution, similar to XRISM/Resolve observations, from the 3-D models and compare them across different explosion scenarios.
By elucidating the differences between models, we demonstrate how the observed diversity of Type Ia SNRs can be reproduced, thereby strengthening the connection between 3-D explosion models and SNRs.
Finally, by comparing our synthetic 3-D models with XRISM observations, we assess whether the explosion models can be qualitatively constrained.

This study is divided into two steps: hydrodynamic simulations and spectral synthesis.
First, we perform a 3-D hydrodynamic simulation as in \citet{ferrand2021supernova,ferrand2025role} and extend it to include time-dependent plasma ionization.
We evolve six Type Ia SNR models in a uniform ISM environment coherently.
These allow us to isolate the effects of explosion properties on SNR evolution.
Next, we synthesize high-spectral resolution X-ray spectra by using these 3-D SNR models.

The structure of this paper is as follows.
Section~\ref{sec:methods} describes the computational framework, including the hydrodynamics code, the 3-D Type Ia SN explosion models, and the overall simulation procedure.
Section~\ref{sec:result} presents the Results and Discussion, which consist of several subsections.
Section~\ref{sec:plasma} presents the analysis of plasma evolution, discussing thermal and bulk properties as well as the elemental abundances and their time evolution.
Section~\ref{sec:X-ray} focuses on X-ray diagnostics, including the Fe-K$\alpha$ emission analysis, the volume-integrated spectra, and spatially resolved spectra that are particularly useful for comparison with XRISM X-ray observations, or high-resolution spectroscopic X-ray observations of future missions like \textit{NewAthena} \citep[e.g.,][]{cruise2025newathena}.
We refer the reader to previous studies \citep[e.g.,][]{ferrand2019supernova,ferrand2021supernova,ferrand2022double,ferrand2025role} for discussions of the global dynamics and morphological evolution of SNRs.

\section{Methods} \label{sec:methods}

\subsection{Code} \label{subsec:code}

We simulate the evolution of SNRs using a 3-D Eulerian hydrodynamic code with NEI treatment based on the PPMLR scheme implemented in VH-1 \citep[e.g.,][]{blondin2001rayleigh}. 
To maintain sufficient resolution throughout the evolution from supernovae to supernova remnants, during which the physical scales change dramatically, we employ a co-moving Cartesian mesh that expands with the SNR. 
The expansion of the SNR is commonly expressed as
\begin{equation}
\label{eq:expansion}
 a(t) = a_{\star}\left(\frac{t}{t_{\star}}\right)^\lambda,
\end{equation}
where $a(t)$ is the SNR size at the time $t$, $a_\star$ is the SNR size at the time $t_{\star}$, and $\lambda$ is the so-called expansion parameter.
We set $t_{\star} = t_0$, the starting time of our simulation.
At each time step, we determine the simulation box required to enclose the expanding SNR and use it to define the scaling factor $a(t)\equiv r_{\text{CD}}(t)/r_{\text{CD}}(t_0)$, where $r_{\text{CD}}$ is the average radius of the contact discontinuity; thus $a(t_0)\equiv 1$.
We then interpolate $a(t)$ (using \texttt{scipy.interpolate.splrep}) to obtain a smooth function, from which we calculate $\lambda(t) = \frac{da/a}{dt/t}=\frac{d\ln (a)}{d\ln (t)}$ and its derivative $\mu(t) = \frac{d\lambda(t)}{d\ln t}$ (using \texttt{scipy.interpolate.spalde}).
Finally, we compute the co-moving time $\tilde{t}$ and perform the hydrodynamic simulations using the coordinate transformation described in \citet{ferrand20103d,ferrand2019supernova}.
%The expansion is described using a scale factor $a(t)$, and the coordinate transformation follows the formulation presented in \cite{ferrand2019supernova}.

Furthermore, we incorporate Lagrangian tracer particles into the Eulerian hydrodynamics code, similar to \citet{seitenzahl2013three}.
We distribute 70,000 particles in the ejecta and 30,000 particles in the ISM.
We confirm that with this number of tracer particles, the results are numerically converged, and increasing the number of particles does not significantly change the derived plasma properties.
While \citet{seitenzahl2013three} sampled particles with equal-mass to calculate nucleosynthesis, we instead adopted equal-volume sampling to achieve better spatial resolution of the ejecta structure in the SNR phase.
In the tracer particle, we solve for the local plasma state that includes $n_{\mathrm{e}}$, $T_{\mathrm{e}}$, abundances, NEI, and $T_{\mathrm{ion}}$, at every hydrodynamical timestep \citep[e.g.,][and references therein]{Court_2024, court2025x}.
This approach ensures that the NEI and related plasma conditions are evolved self-consistently with the hydrodynamics, rather than being treated in a post-processing step.
These quantities are essential for calculating thermal X-ray emission.
For the NEI calculations, we used the ionization and recombination rates from the Astrophysical Plasma Emission Database (APED) of the Atomic Database for Astrophysicists (ATOMDB) \citep[e.g.,][]{smith2005astrophysical}.
The evolution of NEI can be expressed as
\begin{equation}
\begin{split}
 \frac{dn_{i,z}}{dt} = n_{\mathrm{e}} \{ &\alpha_{i,z+1}(T_{\mathrm{e}})n_{i,z+1} + S_{i,z-1}(T_{\mathrm{e}})n_{i,z-1}\\
 &- [\alpha_{i,z}(T_{\mathrm{e}})+S_{i,z}(T_{\mathrm{e}})]n_{i,z} \},
 \end{split}
 \label{eq:nei}
\end{equation}
where $T_{\mathrm{e}}$ is the electron temperature and $n_{i,z}$ is the population of the ion species with atom number $i$, ionization state $z$, $n_{\mathrm{e}}$ is the electron density, $\alpha_{i,z+1}$ is the recombination rate from state $z+1$ to $z$, and $S_{i,z-1}$ is the ionization rate from state $z-1$ to $z$. 
At the shock, we adopt mass-proportional ion heating and set the post-shock electron-to-proton temperature ratio to $T_{\mathrm{e}}/T_{\mathrm{p}} = m_e/m_p$, where $T_{\mathrm{p}}$ is proton temperature. 
This ratio is lower than values typically inferred from observations (e.g., $T_{\mathrm{e}}/T_{\mathrm{p}} > 0.02$, \citet{raymond2023electron}),
but varying the initial ratio has little impact on the NEI evolution of the metal-rich ejecta.
Because the ejecta contain abundant heavy elements, electrons rapidly lose energy to collisional ionization and cool soon after shock passage \citep{yamaguchi2013new}.
In contrast, the assumed $T_{\mathrm{e}}/T_{\mathrm{p}}$ may be more important for the shocked ISM. 
Since the ISM tends to be dominated by the lighter elements, collisional ionization typically removes less energy from electrons, potentially preserving the effect of the initial electron heating in $T_{\mathrm{e}}$.
In any case, our main results, which focus on the ejecta emission, are not sensitive to the assumed initial $T_{\rm e}/T_{\rm p}$.
A more systematic exploration of this dependence is beyond the scope of this paper and will be presented elsewhere.
Subsequently, $T_{\mathrm{e}}$ and $T_{\mathrm{p}}$ evolve toward equilibration via Coulomb collisions at every hydrodynamic timestep.
Using the local temperature and density, we obtained the ionization and recombination rates from tabulated values, and updated the ionization states at each timestep based on the values from the previous timestep to solve the NEI evolution.

The combination of a co-moving grid and tracer particles allows the hydrodynamical evolution and local physical processes to be computed efficiently, which in turn reduces the overall computational cost.
Each 3-D simulation was performed with a spatial resolution of $256^3$ on a workstation using 32 CPU cores in parallel.
The simulations were evolved up to 1000 years, and a single run was completed within about half a day.
This efficiency also allows systematic surveys to be conducted in a wide range of 3-D explosion models.

For the X-ray spectral synthesis, we used the Simulated Observations of X-ray Sources software \citep[SOXS,][]{zuhone2023soxs}, which enables NEI spectral calculations and simulations of observations based on instrument response files for X-ray spectrometers such as XRISM/Resolve.
Using this setup, we evolved the 3-D explosion models described in Section~\ref{subsec:initial} and simulated their SNR evolution as detailed in Section~\ref{subsec:sim}.

\subsection{Initial conditions} \label{subsec:initial}

\begin{deluxetable*}{lcccccccc}
\tablewidth{0pt}
\tablecaption{Model parameters \label{tab:modelparameter}}
\tablehead{
\colhead{\textbf{Model}} & \colhead{Channel} & \colhead{Mechanism} & \colhead{$M_{\text{ej}} \, (M_{\odot})$} &\colhead{$E_{\text{kin}}$ (erg)} & \colhead{$^{56}$Ni} & \colhead{IGE} & \colhead{IME} & \colhead{Reference}
}
\startdata
\texttt{N100ddt}   & \multirow{4}{*}{near-$M_\text{Ch}$} & \multirow{2}{*}{Delayed Detonation} & 1.40 & $1.43 \times 10^{51}$ & 43\% & 60\% & 32\% & \multirow{2}{*}{\citet{seitenzahl2013three}} \\
\texttt{N5ddt}     &                     &                                      & 1.40 & $1.55 \times 10^{51}$ & 70\% & 81\% & 14\% &  \\
\cline{3-3}
\cline{9-9}
\texttt{N100def}   &                     & \multirow{2}{*}{Deflagration}        & 1.31 & $6.15 \times 10^{50}$ & 27\% & 42\% & 11\% & \multirow{2}{*}{\citet{fink2014three}} \\
\texttt{N5def}     &                     &                                      & 0.372 & $1.35 \times 10^{50}$ & 42\% & 60\% & 11\% &  \\
\cline{2-3}
\cline{9-9}
\texttt{OneExp}  & \multirow{2}{*}{sub-$M_\text{Ch}$} & \multirow{2}{*}{Double Detonation}   & 1.09 & $1.4 \times 10^{51}$  & 41\% & 46\% & 39\% & \multirow{2}{*}{\citet{pakmor2022fate}} \\
\texttt{TwoExp}  &                     &                                      & 1.75 & $1.9 \times 10^{51}$  & 26\% & 29\% & 42\% &  \\
\enddata
\tablecomments{Model parameters for 6 explosion models. $M_{\text{ej}}$ represents the ejected mass in solar mass units, and $E_{\text{kin}}$ represents the kinetic energy in ergs. The columns for $^{56}$Ni, iron-group elements (IGE, including $^{56}$Ni), and intermediate-mass element (IME) list the mass fractions relative to the total ejecta mass. More detailed discussions on each model are provided in the corresponding references.}
\end{deluxetable*}

\begin{figure}[ht!]
\includegraphics[width=\columnwidth]
{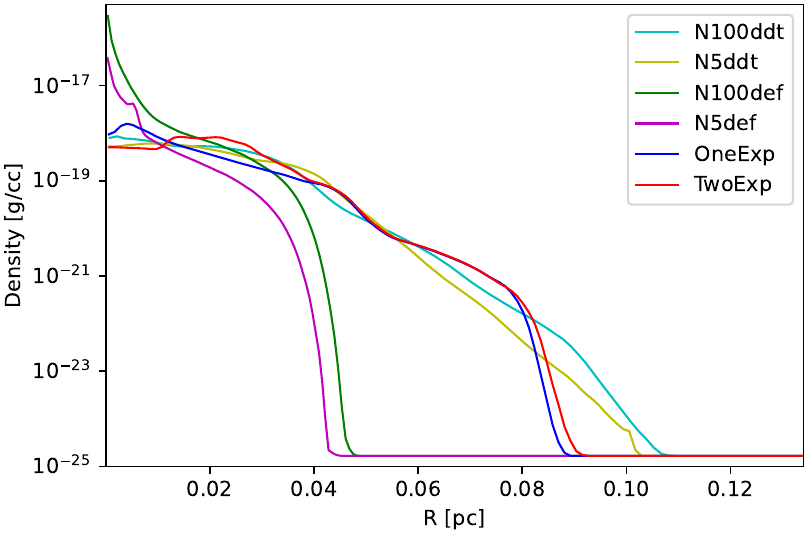}
\caption{For illustration, we show 1-D radial profiles of the initial density at 3 years, which were obtained for six Type Ia SN explosion models by spherically averaging the 3-D initial conditions. Although the simulations are fully 3-D, this 1-D representation provides a clearer view of the radial structure. Up to 3 years, SNRs are assumed to expand freely.
\label{fig:density_init}}
\end{figure}

\begin{figure}[ht!]
\includegraphics[width=\columnwidth]{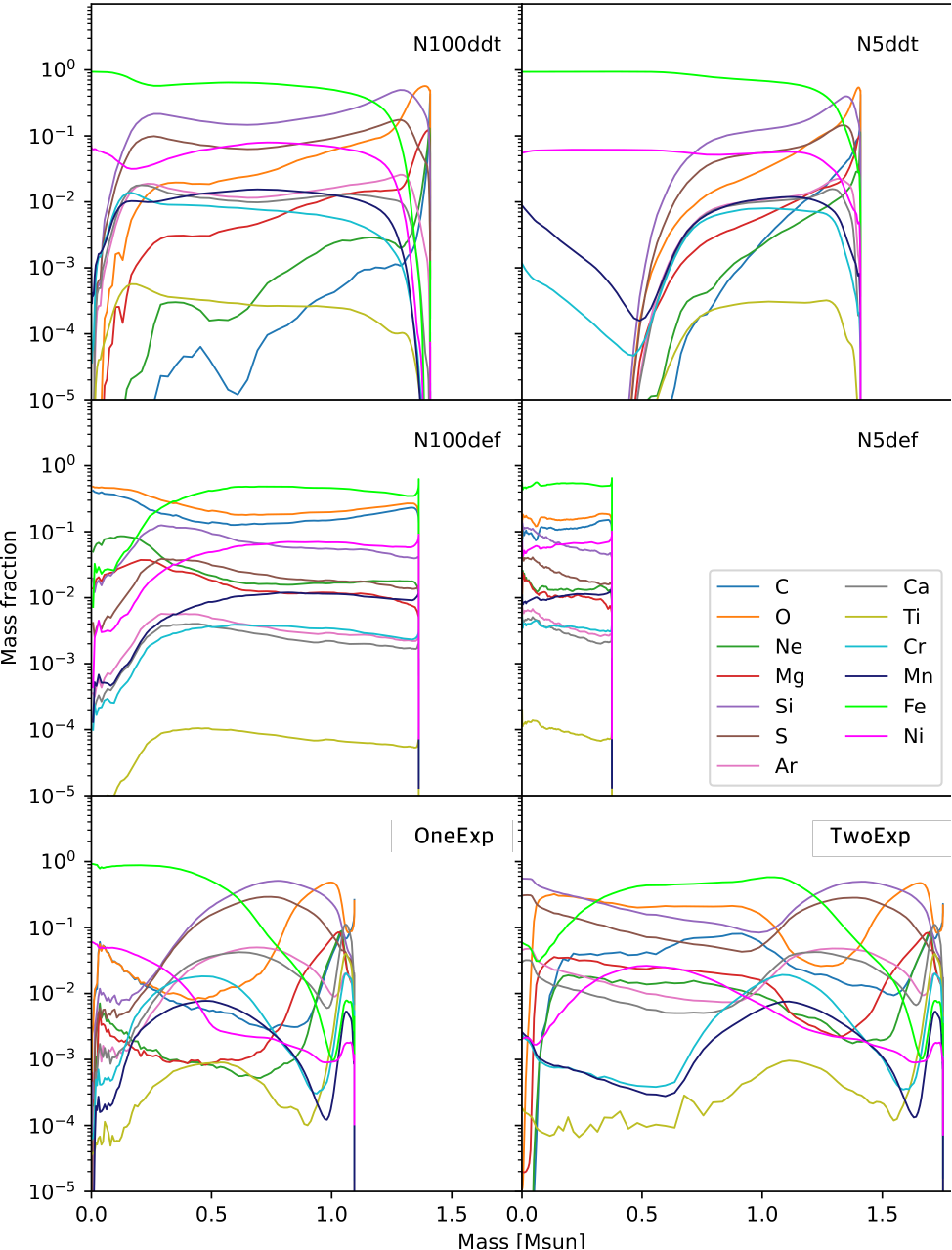}
\caption{Initial abundance distribution for each model. As in Figure~\ref{fig:density_init}, the 3-D data were spherically averaged into a 1-D radial profile.
\label{fig:abund_init}}
\end{figure}

We evolved six Type Ia SN explosion models, selecting two from each of the literature \citet{seitenzahl2013three}, \citet{fink2014three}, and \citet{pakmor2022fate}. 
The models from \citet{seitenzahl2013three} and \citet{fink2014three} are identical to those analyzed in \citet{ferrand2021supernova}, while the two models from \citet{pakmor2022fate} correspond to those in \citet{ferrand2025role}. 
Of these, \texttt{N100ddt}, \texttt{N5ddt}, \texttt{N100def}, and \texttt{N5def} are near-$M_\text{Ch}$ models; \texttt{OneExp} and \texttt{TwoExp} are sub-$M_\text{Ch}$ models. 
Their main parameters are summarized in Table~\ref{tab:modelparameter}.
For clarity, the near-$M_\text{Ch}$ models considered here do not include an explicit companion (following the original works), whereas the sub-$M_\text{Ch}$ models include a secondary white dwarf. 
This distinction is relevant for interpreting asymmetries, because the presence of a companion can imprint directional structure on the ejecta.

In Figure~\ref{fig:density_init} and Figure~\ref{fig:abund_init}, the initial 3-D distribution (density and abundance, respectively) is spherically averaged to produce a 1-D profile. 
Although the 3-D distribution is neglected in these figures, they are useful for a simplified understanding of the explosion properties.

First, we provide an overview of the near-$M_\text{Ch}$ models.
`ddt' stands for deflagration-to-detonation transition (or delayed detonation, as indicated in Table~\ref{tab:modelparameter}), and \texttt{N100ddt} and \texttt{N5ddt} are the models in \citet{seitenzahl2013three}.
`def' stands for deflagration, and \texttt{N100def} and \texttt{N5def} are the models in \citet{fink2014three}.
These numbers indicate the number of ignition points. These values affect the asymmetry of the explosion.

The \texttt{N100ddt} model, a delayed detonation model, assumes a nearly spherical geometry.
A large number of ignition points leads to a strong deflagration phase, during which the WD expands significantly before the detonation sets in.
As a result, the ejecta density becomes relatively low, and a large amount of intermediate-mass elements (IMEs) are synthesized during the detonation phase.
In contrast, the \texttt{N5ddt} model, despite being the same type of delayed detonation model, exhibits strong asymmetry due to the small number of ignition points.
The deflagration phase is weak, and the WD does not expand significantly.
Consequently, the ejecta retain a relatively high density during the detonation phase, leading to the synthesis of a large amount of iron-group elements (IGEs), particularly $^{56}$Ni.
This results in a relatively bright supernova.
In \citet{Stone_2021}, 2-D and 3-D hydrodynamic simulations were performed using the \texttt{N3} model from \citet{seitenzahl2013three}, a delayed detonation model with even stronger asymmetry than \texttt{N5ddt}, to reproduce the strong asymmetry observed in the youngest known Type Ia SNR, G1.9+0.3.
Although the model does not fully reproduce the observed remnant, the simulations exhibit asymmetric structures and elemental distributions, highlighting the need to investigate asymmetric explosion models such as \texttt{N3} and \texttt{N5ddt}.

The \texttt{N100def} model is a pure deflagration model that exhibits an approximately spherical geometry due to the large number of ignition points.
Because of the large ignition point number, it produces a relatively strong explosion compared to other deflagration model, in which explosions are generally weak.
As a result, almost no bound remnant of the WD remains.
In this model, only a small amount of IMEs are synthesized during the deflagration wave.
No known subclass of observed Type Ia supernovae matches this model well.
In contrast, the \texttt{N5def} model is also a pure deflagration model but has a much smaller number of ignition points, resulting in a weaker explosion.
Consequently, a bound WD remnant with a mass of 1.03 $M_{\odot}$ remains at the center.
As with \texttt{N100def}, only a small amount of IMEs are produced.
This model has been suggested as a candidate for explaining faint Type Ia supernovae such as SN 2002cx-like events.

Next, we describe the remaining two sub-$M_\text{Ch}$ models, which are based on the DD scenario \citep{pakmor2022fate}.
The progenitor binary systems in these models consist of two carbon–oxygen WDs with thin helium shell.
“One” and “Two” in the model names indicate the number of WDs that explode.
In both models, helium mass transfer occurs from the secondary to the primary WD, leading to the ignition of helium detonation in the thin helium shell on the surface of the primary.
The shock wave of the helium detonation propagates inward and reaches the carbon core, where it triggers a carbon detonation.
In this way, a successful explosion is achieved via the double detonation mechanism.
In the \texttt{OneExp} model, the secondary WD does not explode, whereas in the \texttt{TwoExp} model, the secondary also explodes through the same double detonation mechanism.
This secondary explosion is initiated manually.
In 3-D radiative-transfer models corresponding to our \texttt{OneExp} and \texttt{TwoExp}, \citet{pollin2024fate} showed that both configurations can reproduce observables of 02es-like SNe Ia \citep[e.g.,][]{ganeshalingam2012low}, which are considered peculiar. 
The results exhibit a strong viewing-angle dependence. 
Moreover, when the secondary detonates and helium-ash signatures are weak, the spectra can approach those of normal SNe Ia.

\subsection{Simulations} \label{subsec:sim}

We performed 3-D simulations of the six explosion models in a uniform-density ISM environment, evolving them from 3 to 1000 years.
Up to 3 years, the SNRs were assumed to expand freely.

The ISM number density was set to $n_{\text{ISM}} = 0.3\,\mathrm{cm}^{-3}$ corresponding to mass density $\rho_\mathrm{ISM}\sim0.5\times10^{-24} \,\mathrm{g}\,\mathrm{cm}^{-3}$, typical for Tycho's SNR \citep{williams2013azimuthal, slane2014cr}, and the ISM temperature was set to $T_{\text{ISM}} = 10^4 \,\mathrm{K}$. 
The ISM abundances were assumed to follow the  solar abundances of \citet{asplund2009chemical}.
The output of the hydrodynamic simulation includes position, velocity, density, and pressure.
For the tracer particles, we recorded only shocked particles including position, velocity, mass, $T_{\mathrm{ion}}$, $n_{\mathrm{ion}}$, $T_{\mathrm{e}}$, $n_{\mathrm{e}}$, NEI and abundances.

Subsequently, we computed the X-ray emission using the tracer particles at 100, 400, 700, and 1000 years.
The energy resolution was set to approximately 1 eV, which is higher than the resolution of XRISM/Resolve instrument ($\sim$4.5 eV), to retain fine spectral features while enabling direct comparison with XRISM observation data.
For absorption, we adopted the \texttt{tbabs} model, which accounts for interstellar medium absorption \citep{wilms2000absorption}.
We assumed an SNR distance of 2.5\,kpc and a column density of $0.6 \times 10^{22}\,\mathrm{H\,cm^{-2}}$, corresponding to values estimated for Tycho's SNR \citep[e.g.,][]{badenes2006constraints}.
The output of the X-ray simulation consists of one spectrum per shocked tracer particle.
In this post-processing approach, the spectra from individual tracer particles are integrated to obtain the total spectrum.
We neglect absorption and scattering within the remnant and apply only a foreground ISM absorption model in the spectral synthesis, since the remnant is optically thin to its own X-ray emission.
As an order-of-magnitude upper limit, we take the post-shock density to be $n_\mathrm{H} \simeq 4\,n_{\rm ISM}$, corresponding to the canonical compression ratio for a strong shock in an ideal monatomic gas ($\gamma=5/3$), and adopt a length scale $L$.
We can place an upper limit on internal photoelectric absorption by adopting a neutral-gas effective cross section per H nucleus, $\tau_{\rm pe}(E)=N_{\rm H}\sigma_{\rm pe,eff}(E)$ \citep{morrison1983interstellar}.
Conservatively taking $N_{\rm H}\simeq 4\,n_{\rm ISM}L$ and $\sigma_{\rm pe,eff}(0.3~{\rm keV})\approx 3\times10^{-21}\,{\rm cm^2}$ yields
\begin{equation}
\tau_{\rm pe}(0.3~{\rm keV}) \lesssim 0.1\left(\frac{n_{\rm ISM}}{0.3~{\rm cm^{-3}}}\right)\left(\frac{L}{10~{\rm pc}}\right),
\end{equation}
and $\tau_{\rm pe}$ decreases rapidly with increasing energy.
Since the shocked plasma is highly ionized, using neutral-gas cross sections provides a conservative upper limit on any internal photoelectric attenuation.
We also estimate the Thomson optical depth as
\begin{equation}
 \tau_{\rm T} = N_e\sigma_{\rm T} \sim 2.5\times10^{-5}\left(\frac{n_\mathrm{ISM}}{0.3\,\mathrm{cm}^{-3}}\right)\left(\frac{L}{10\,\mathrm{pc}}\right),
\end{equation}
where $N_e$ is electron column density and $\sigma_{\rm T}=6.65\times10^{-25}\,{\rm cm^2}$ is the Thomson cross section.
This method also enables the calculation of local spectra and the Doppler effects based on the kinematic information of the emitters along specific line-of-sight.

\section{Results and Discussion} \label{sec:result}

\subsection{Plasma evolution} \label{sec:plasma}

\subsubsection{Thermal properties}

\begin{figure*}[ht!]
\centering
\includegraphics[width=0.85\textwidth]{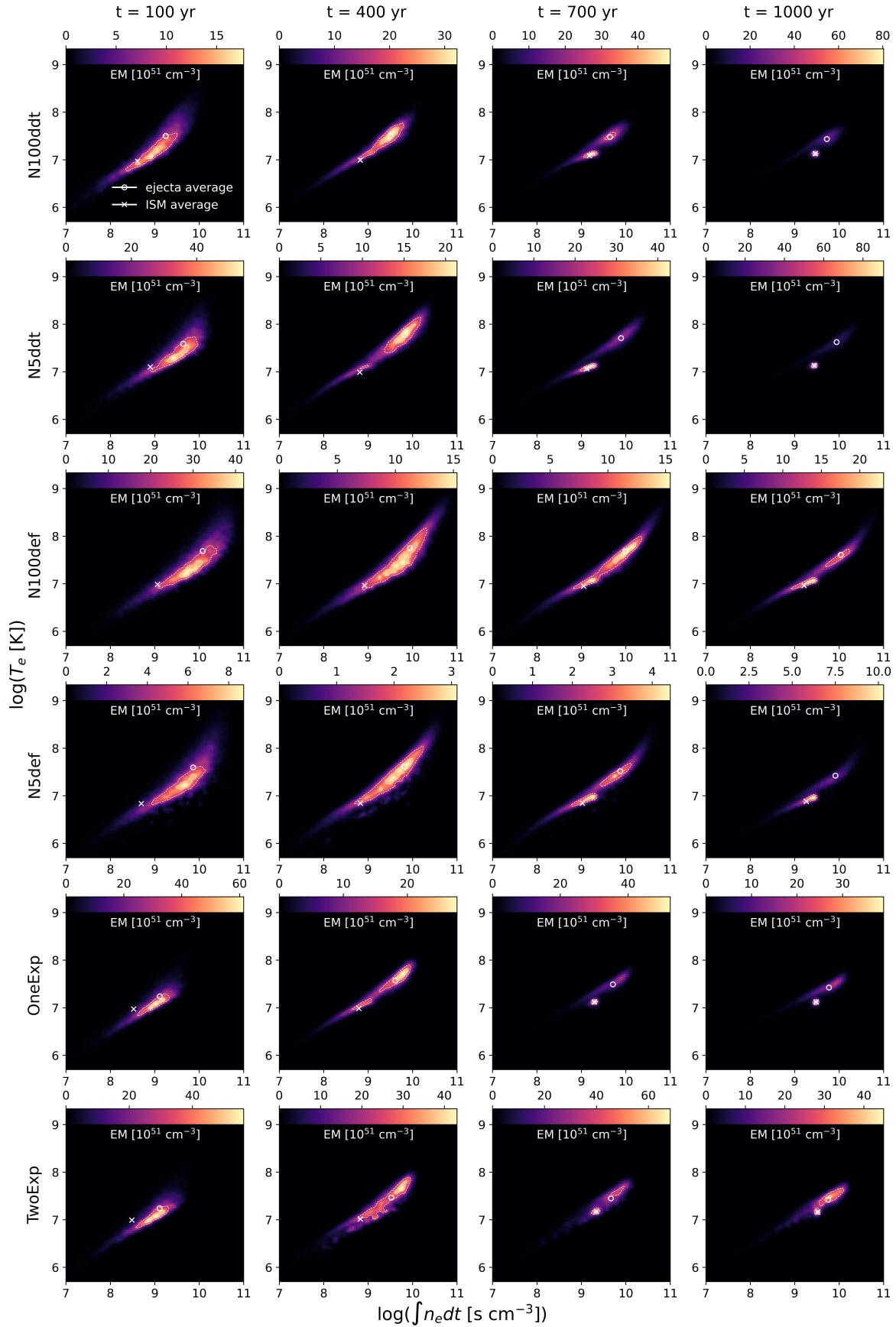}
\caption{The distribution of emission measure (EM) of IME and IGE elements in the shocked plasma over the $\int n_{\mathrm{e}}  dt$–$T_{\mathrm{e}}$ phase space, where $\int n_{\mathrm{e}}  dt$ is the ionization time.  
Panels from left to right correspond to different evolutionary stages at 100, 400, 700, and 1000 years after explosion. Each row corresponds to the SNR from a particular explosion model. At each epoch, the solid and dashed contours indicate the 80\% and 50\% levels of the peak EM respectively, while the circle (cross) marker denotes the EM-weighted average plasma state in the shocked ejecta (ISM).
\label{fig:net}}
\end{figure*}

\begin{figure*}[ht!]
\includegraphics[width=0.99\textwidth]{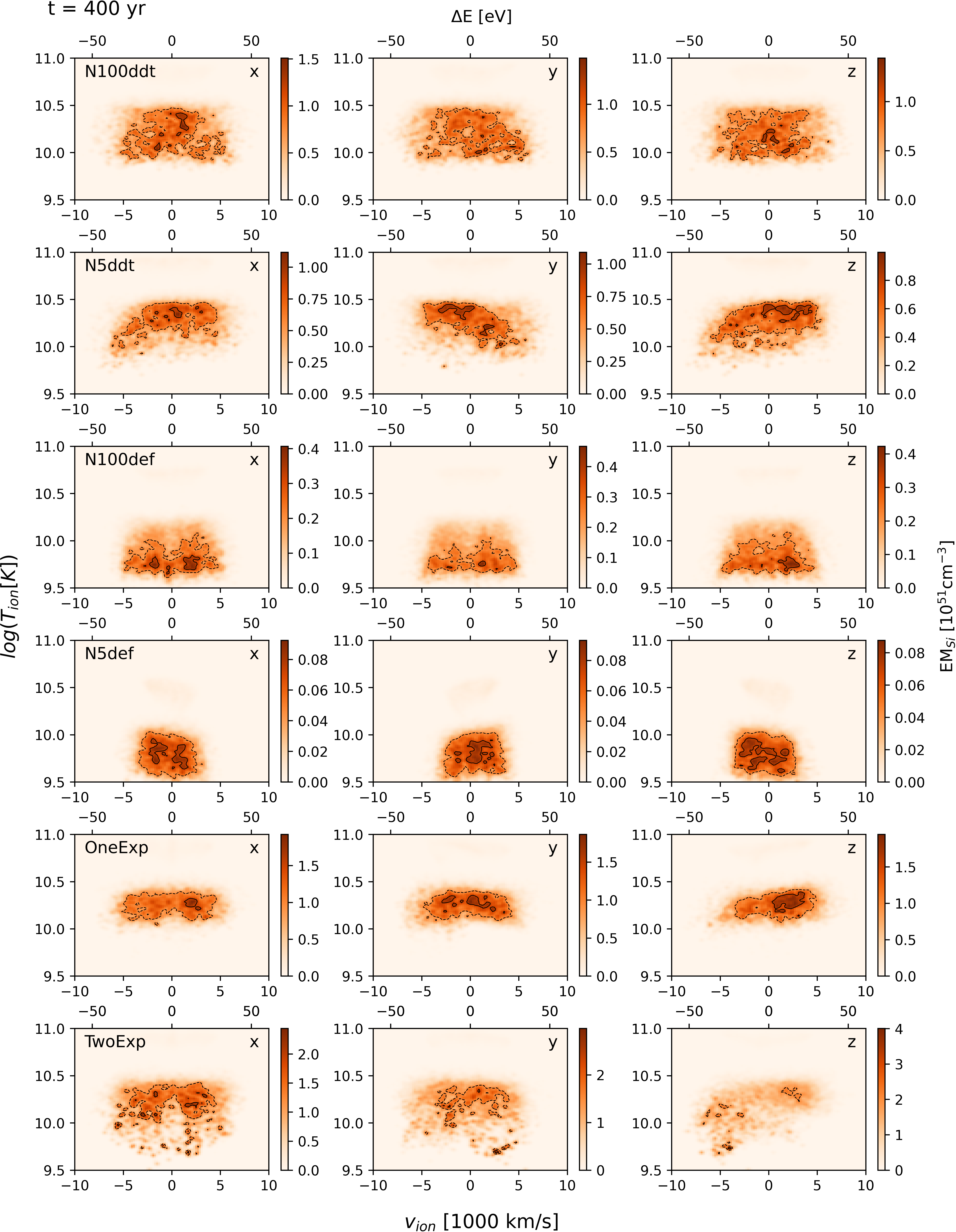}
\caption{The distribution of EM from Si in the shocked plasma over the $T_{ion}$-velocity phase space at 400 years. Here, $v_{\mathrm{ion}}$ is the projected velocity of Si along three orthogonal line-of-sights in the X, Y and Z directions respectively from the left to right panels, which shows the effect of Doppler shift on the emission lines and its dependence on the observing angle. $T_{\mathrm{ion}}$ is the temperature of Si in the shocked plasma which links directly to the level of thermal line broadening.
The upper axis shows the Doppler shift corresponding to the velocity, calculated as $\Delta E = v/c \,\times\,E_{0,Si}$. Here, we adopt $E_{0, Si} = 1830\, \textrm{eV}$, corresponding to the centroid energy of the Si K$\alpha$ line after applying thermal broadening in the \texttt{N100ddt} model.
\label{fig:T_v_Si_400}}
\end{figure*}

\begin{figure*}[ht!]
\includegraphics[width=0.99\textwidth]{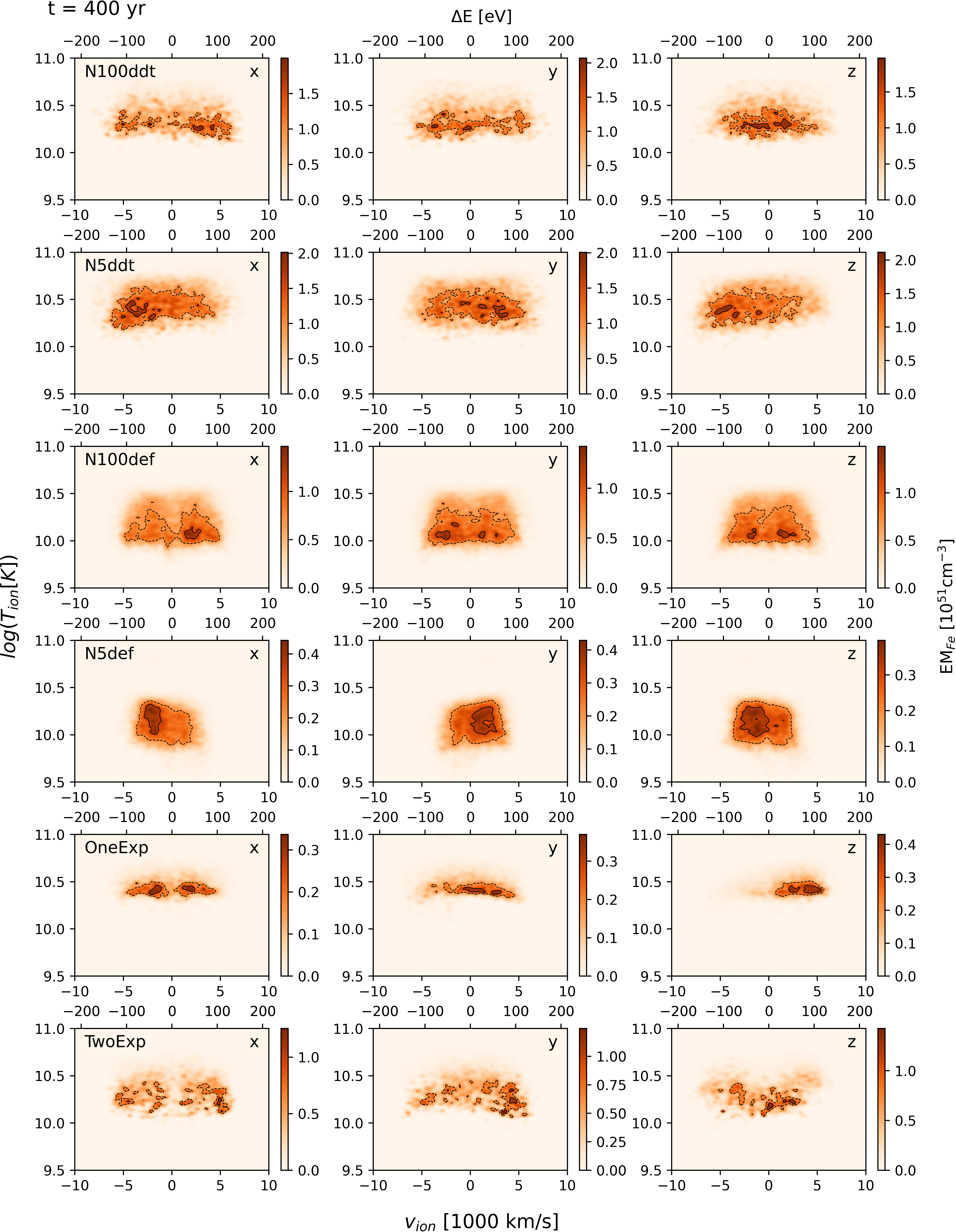}
\caption{Same as Figure~\ref{fig:T_v_Si_400} but for the element Fe.
Here, we adopt $E_{0, Fe} = 6422\, \textrm{eV}$, corresponding to the centroid energy of Fe-K$\alpha$ line after applying thermal broadening in the \texttt{N100ddt} model.
\label{fig:T_v_Fe_400}}
\end{figure*}

The characteristics of X-rays are discussed in this section using the plasma state.

Figure~\ref{fig:net} shows the ionization time and electron temperature of the shocked material weighted by the emission measure EM.
The emission measure, expressed as 
\begin{equation}
EM_{\mathrm{ion}} = \int n_{\mathrm{e}} n_{\mathrm{ion}} dV
\label{eq:EM}
\end{equation}
serves as an indicator of the intensity of X-ray emission.
Here, $n_{\mathrm{e}}$ represents the electron density and $n_{\mathrm{ion}}$ represents the ion density, defined for a single element or as the total over several elements. In Figure~\ref{fig:net}, the EM is defined for oxygen to nickel combined.

This diagram illustrates the time evolution of each model.
The distribution shifts toward higher ionization times as the SNR evolves, reflecting the gradual increase in $\int n_{\mathrm{e}} dt$.
The overall ranges of the thermodynamic properties are similar and broadly consistent with observations of Type Ia SNRs \citep[e.g.,][]{sawada2019ionization,godinaud2025bayesian}.

At early epochs, the plasma states exhibit a broad dispersion but progressively concentrate around the EM-weighted averages (circle and cross markers) at 700 and 1000 years.
Meanwhile, the contribution of shocked ISM gradually increases. 
Although the emission is dominated by shocked ejecta in the first few hundred years, the shocked ISM component becomes more prominent at later stages, eventually dominating the total X-ray emission. 
In ISM environments, the swept-up mass increases as $M_{\text{swept}} \propto r_{\text{FS}}^{3}$, where $r_{\text{FS}}$ is the radius of forward shock, so the switch in dominance from the ejecta to the ISM emission occurs as the remnant expands.

The timing of this transition depends on the explosion model. 
In the deflagration models, the slower expansion velocity reduces the growth of swept-up ISM, allowing the ejecta to dominate the X-ray emission for longer time than in other models. 
In contrast, the delayed-detonation models expand more rapidly, so the swept-up ISM grows faster and overtakes the ejecta contribution earlier. 
Even in the energetic \texttt{N100ddt} case, the ISM becomes dominant at late epochs, although the ejecta retain a more substantial contribution than in weaker models. 
These differences indicate that the relative weight of ejecta versus ISM emission provides a sensitive diagnostic of the explosion mechanism.

At any given epoch, the distributions also show diversity in ionization timescale among the models. 
For example, at 100 years the \texttt{N5ddt} and deflagration models are shifted toward the right side of the diagram compared to the \texttt{N100ddt} model, indicating larger ionization timescales. 
This trend arises because the weaker initial deflagration leaves higher-density material in the inner ejecta, enabling ionization to proceed more efficiently once these regions are shocked. 
Thus, even when the overall EM distribution converges at later times, the early-phase ionization history retains a clear dependence on the explosion mechanism.

Figure~\ref{fig:T_v_Si_400} shows the distribution of EM from Si in the shocked plasma over the $T_{\mathrm{ion}}$--velocity space at 400 years.
Silicon is a characteristic nucleosynthetic product of Type Ia supernovae, and its emission lines are clearly detected in X-ray observations. 
We therefore use Si as a representative tracer to assess the constrainability of line-of-sight velocities. 
Furthermore, as shown later in Figure~\ref{fig:T_v_Fe_400}, we perform the same analysis for Fe, allowing us to evaluate element-to-element differences arising from their 3-D distributions.

Clear differences are found among the models in both ion temperature and velocity. 
The deflagration models are distributed in comparatively lower temperature regions, which can be attributed to their smaller $E_{\mathrm{kin}}$ values (Table~\ref{tab:modelparameter}), resulting in weaker shock heating. 
Their velocity range is also narrower, consistent with the slower expansion velocity expected from the combination of $M_{\mathrm{ej}}$ and $E_{\mathrm{kin}}$.

The contour levels highlight additional differences. At the 50 \% level (dashed), the \texttt{N100ddt}, \texttt{N5ddt}, and \texttt{TwoExp} models exhibit patchy distributions, whereas the other models are largely enclosed within a single coherent contour. 
At the 80 \% level (solid), \texttt{N100ddt} and \texttt{TwoExp} display the most dispersed structures, while \texttt{N5ddt} and \texttt{OneExp} show compact but systematically offset regions in velocity space.
In terms of overall morphology, the near-$M_\text{Ch}$ \texttt{N100} model remains nearly symmetric, whereas the other models present tilted structures. 
These features reflect the spatial distribution of Si: under spherical symmetry, one expects a central, symmetric concentration, while explosion asymmetries introduce lateral variations in velocity and corresponding temperature anisotropies.

The observational implications can be assessed through Doppler line shifts and thermal broadening.
The Doppler shift in the non-relativistic case is given by 
\begin{equation}
\Delta E=\frac{v}{c} E_0
\label{eq:Doppler}
\end{equation}
where $E$ is the observed line energy, $E_0$ is the line energy in the rest frame, 
$v$ is the velocity of the emitter towards the observer, and 
$c$ is the speed of light.
The standard deviation of a Gaussian line profile due to thermal broadening is given by
\begin{equation}
\sigma=\sqrt{\frac{k_{\mathrm{B}} T_{\mathrm{ion}}}{mc^2}}E_0
\label{eq:sigma_thermal}
\end{equation}
where $k_{\mathrm{B}}$ is the Boltzmann constant, $T_\mathrm{ion}$ is the ion temperature, $m$ is the mass of the emitting particle, and $c$ is the speed of light.
Regarding the Si-K$\alpha$ line, its centroid energy is around 1.8\,keV. For a line-of-sight velocity of 5000\,km\,s$^{-1}$, the energy shift is estimated as $\Delta E \sim 30$\,eV. 
For thermal broadening, the standard deviation is given by $\sigma = \sqrt{\frac{k_\mathrm{B} T_\mathrm{ion}}{mc^2}}E$, which yields $\sigma \sim 10$\,eV at $T = 10^{10}$\,K.

Thus, while the asymmetries in Si distributions are evident in Figure~\ref{fig:T_v_Si_400}, the peak ion temperatures ($\sim 10^{10}$ K) and line-of-sight velocities (2000–3000 km s$^{-1}$) imply that thermal broadening is comparable to or larger than Doppler broadening at 400 years. Consequently, the spectral imprint of Doppler shifts is expected to be relatively minor in integrated spectra, even though the underlying ejecta asymmetries remain significant.

Figure~\ref{fig:T_v_Fe_400} is same as Figure~\ref{fig:T_v_Si_400} but for the element Fe.
Overall, the degree of scatter shows features similar to those of Si, and the degree of asymmetry is expected to follow the same trend.
However, in the case of Fe, the temperature range is narrower than that of Si, and the direction of the offset often differs from that of Si. 
For example, focusing on the 80\% contour of \texttt{N5ddt}, the Si distribution appears to be shifted toward (x: center, y: minus, z: plus), whereas the Fe distribution appears to be shifted toward (x: minus, y: plus, z: minus). 
This indicates that the spatial distributions of Si and Fe are not identical. 
Nevertheless, in models with strong asymmetries, such as \texttt{OneExp}, the trend of being shifted toward positive z is still preserved.

For the Fe-K$\alpha$ line, the observational implications differ substantially from those of Si. 
Its centroid energy is around 6.4\,keV, a line-of-sight velocity of 5000\,km\,s$^{-1}$ corresponds to an energy shift of $\Delta E \sim 110$\,eV. 
The thermal broadening at $T = 10^{10}$\,K is only $\sigma \sim 20$\,eV, significantly smaller than the Doppler shift. 
As shown in Figure~\ref{fig:T_v_Fe_400}, the Fe distributions tend to exhibit more pronounced asymmetries than those of Si, with the \texttt{N5ddt} and \texttt{TwoExp} models containing components at velocities approaching 5000 km s$^{-1}$. 
Consequently, the Doppler effect is expected to dominate over thermal broadening in Fe, making the asymmetries in its distribution more directly visible in the observed spectra.

\subsubsection{Abundance ratio}

\begin{figure*}[ht!]
\includegraphics[width=\textwidth]{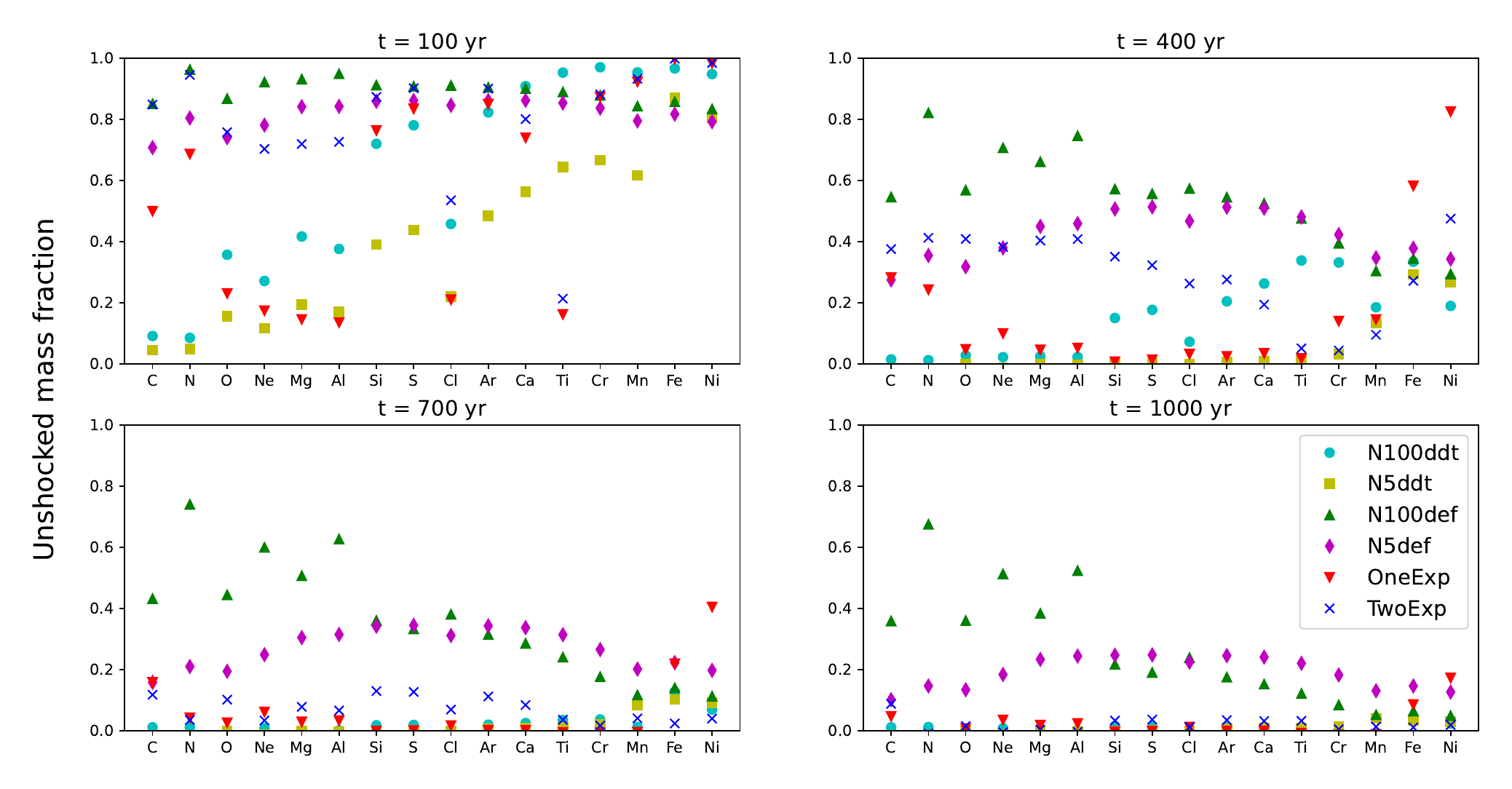}
\caption{Unshocked mass fractions of several important elements in the ejecta for each model. 
From top left to bottom right, the panels correspond to 100, 400, 700, and 1000 years, respectively.
The unshocked mass fraction of element $i$ is defined as $M_{i,\mathrm{unshocked}}/(M_{i,\mathrm{shocked}}+M_{i,\mathrm{unshocked}})$ within the ejecta.}
\label{fig:unshocked_mass_frac}
\end{figure*}

\begin{figure*}[ht!]
\includegraphics[width=\textwidth]{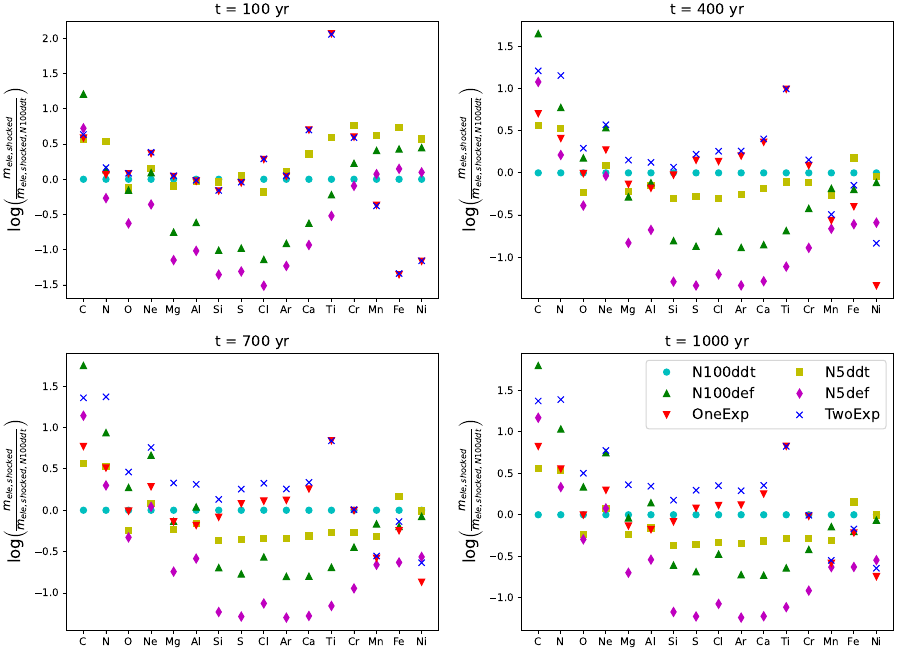}
\caption{Shocked ejecta masses of the same elements as in Figure 6 for each model, normalized by the shocked mass in the \texttt{N100ddt} model. From top left to bottom right, the panels correspond to 100, 400, 700, and 1000 years, respectively.
\label{fig:abund_pattern_diff}}
\end{figure*}

Figure~\ref{fig:unshocked_mass_frac} shows the unshocked mass fraction of several important elements in the ejecta.
Here, the unshocked mass fraction of element $i$ is defined as $M_{i,\mathrm{unshocked}}/(M_{i,\mathrm{shocked}}+M_{i,\mathrm{unshocked}})$ within the ejecta.
Therefore, element by element, a smaller value indicates that a larger fraction of that element has been shocked.
Not only among the models but also among the elements, the evolution of the shocked mass fraction differs.
These can be understood from the initial abundance distributions shown in Figure~\ref{fig:abund_init}.
Because the reverse shock propagates inward from the outer layers, elements concentrated in the outer ejecta are shocked earlier, while those located deeper remain unshocked until later times.

Focusing on the delayed detonation models in Figure~\ref{fig:abund_init}, light elements such as carbon and oxygen have high mass fractions in the outer ejecta and low fractions in the inner regions.
As a result, their unshocked fractions become small already at 100 years.
In contrast, heavier elements such as Fe and Ni are concentrated in the central regions, leading to high unshocked fractions at 100 years that gradually decrease as the reverse shock penetrates inward.

The deflagration models show distinct behavior compared to the delayed detonation models.
Because the elemental composition is relatively homogeneous regardless of radius, the unshocked mass fractions of all elements exhibit similar evolutionary behavior.
However, in the \texttt{N100def} model, the mass fractions of carbon and oxygen are higher near the center, which explains why their unshocked mass fractions remain relatively high even after 400 years.

The DD models feature a more complex, 3-D elemental distribution.
The rapid decrease in the unshocked mass fractions of elements such as Mg and Ti can be explained by the distribution of abundances shown in Figure~\ref{fig:abund_init}.

The unshocked mass fraction is an essential quantity for constraining models of young SNRs. 
Not all of the ejecta are shocked, and the 3-D distribution of elements introduces biases into the observed line ratios. 
Spectral fitting has often assumed that the observed abundances directly reflect the explosion yields, but this approach neglects the effects of shock evolution. 
Our results demonstrate that line ratios should instead be interpreted by accounting for the unshocked mass fractions derived from simulations, ensuring a more accurate connection between explosion models and observed spectra.

Figure~\ref{fig:abund_pattern_diff} shows the shocked ejecta masses normalized by the shocked mass in the \texttt{N100ddt} model.
This comparison clearly highlights differences in elemental composition and distribution across the explosion models, with the overall characteristics largely separated by explosion mechanism.

Overall, the characteristics of the shocked masses are separated according to the explosion mechanism.
Taking the \texttt{N100ddt} model as the reference, the \texttt{N5ddt} model is distributed close to it.
As described in Section~\ref{subsec:initial}, the \texttt{N5ddt} model contains fewer IMEs and more IGEs compared to the \texttt{N100ddt} model.
At 100 years, however, some of the lighter elements show values comparable to or even exceeding those of \texttt{N100ddt}, an inversion that reflects differences in the radial distribution of abundances. 
By 400 years, the two models converge toward similar trends.
The deflagration models are located at lower values because of their smaller ejecta mass. 
As shown in Table~\ref{tab:modelparameter}, their composition contains only about 10 \% IMEs, resulting in a lower central distribution.
The DD models exhibit shocked masses that differ significantly from those of the \texttt{N100ddt} model. 
In particular, their higher Ti and lower Ni abundances are remarkable and represent characteristic features of this class. 
Moreover, they tend to have fewer IGEs and more IMEs, consistent with the smaller mass of the exploding WD.

Since shocked masses are directly related to line luminosities, these differences are expected to manifest in observed spectra. 
The temporal inversion of shocked masses suggests that the apparent composition of the shocked ejecta may evolve with remnant age.
While this effect cannot be followed observationally on hundred-year timescales, it provides a useful indication that the shocked abundances do not simply mirror the total explosion yields, and may therefore serve as one of several clues when interpreting observed line ratios.

\subsection{X-ray properties}\label{sec:X-ray}

\subsubsection{Iron K$\alpha$ line diagnostics} \label{sec:yamaguchi}

\begin{figure*}[ht!]
\includegraphics[width=\textwidth]{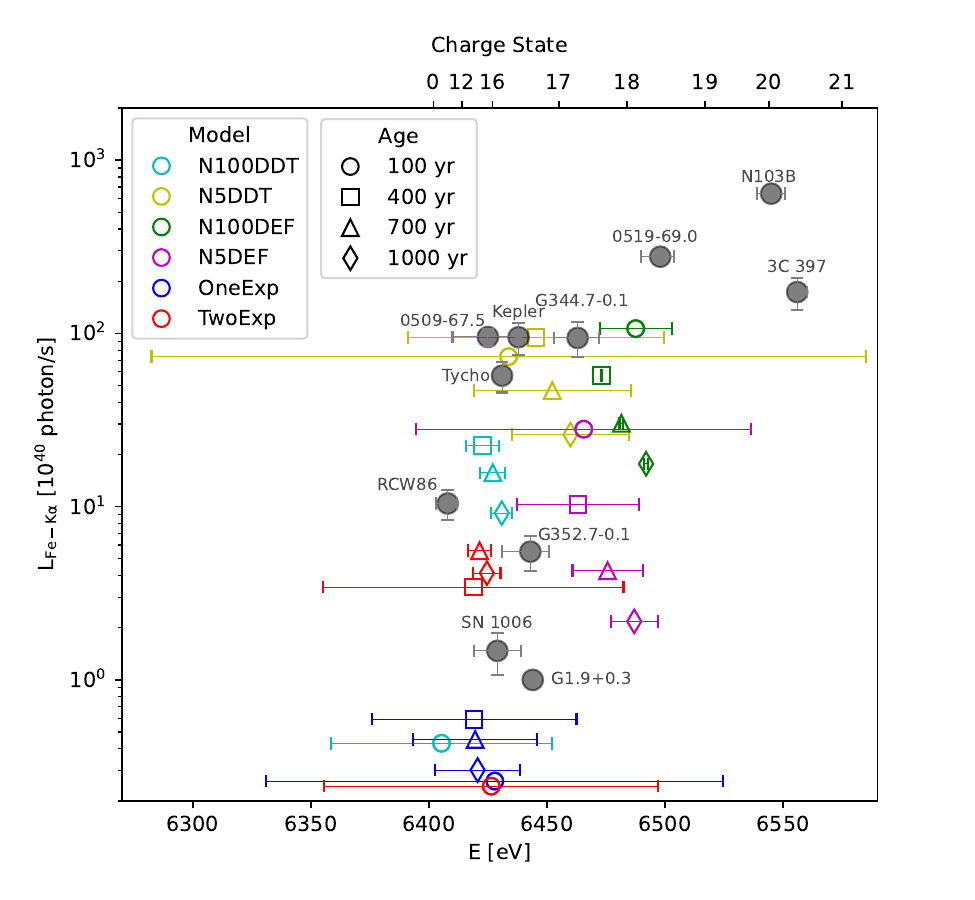}
\caption{The centroid energy and line luminosity of Fe-K$\alpha$ for the six SNR models at four different ages, in comparison with Type Ia SNR observations \citep{yamaguchi2014discriminating}. The colored symbols are the estimated values from the simulations, without the effect of Doppler shifts from bulk motion. Their errorbars show the maximum possible effect from Doppler shift over all possible viewing angles, which evolves with time and differ among the models according to the expansion speed and different degrees of initial explosion asymmetry.  
\label{fig:Yamaguchi}}
\end{figure*}

Figure~\ref{fig:Yamaguchi} shows the centroid energy and line luminosity of Fe-K$\alpha$ line for the six SNR models at four different epochs, in comparison with Type Ia SNR observations.
These quantities were calculated using only the Fe-K$\alpha$ line components, excluding the continuum. 
The line luminosity is obtained by summing the luminosity in each energy bin, while the centroid energy is computed as the luminosity-weighted average energy across the line profile.
The error bars indicate the maximum Doppler shift due to the line-of-sight direction. 
They were calculated using the viewing angle that gives the maximum absolute value of the EM-weighted projected velocity of Fe in the shocked ejecta, as referenced in Appendix~\ref{sec:append_dist_vlos}.
At younger ages, the explosion asymmetry remains more prominent. 
In asymmetric models, the Doppler shift is significant, indicating that the interpretation of the Fe-K$\alpha$ line centroid energy requires not only the ionization state but also the effect of explosion asymmetry.

The centroid energy of the Fe-K$\alpha$ line corresponds to the charge state of iron.
This line is emitted when an electron transitions from the L-shell to the K-shell. 
When the outer shells are filled, the electrons in the K and L shells experience a reduced effective nuclear charge due to the screening effect of the outer electrons. As a result, the energy difference between the K and L shells decreases, leading to a lower photon energy for the transition.

The simulation results in this study generally reproduce the diversity observed in Type Ia supernova remnants, primarily reflecting differences in explosion models. 
While this study assumes an ISM density of $0.3\,\text{cm}^{-3}$, small shifts in centroid energy and line luminosity can occur depending on the ambient density, with higher densities yielding higher values.
The simulation results show that, with the exception of the \texttt{N100ddt} model at 100 years, the models can be broadly classified into three regions according to their explosion mechanisms. 
The first group comprises the DDT models, which are located in the region of relatively low ionization state and relatively high line luminosity. 
The second group includes the deflagration (DEF) models, which exhibit relatively high ionization states and also relatively high line luminosities. 
The third group consists of the DD models, which lie in the region of both low ionization and low line luminosity.
Each of these regions corresponds to certain observed SNRs, suggesting that Fe-K$\alpha$ line diagnostics can serve as a powerful tool to constrain explosion models. 

However, some remnants, such as 0519–69.0, N103B, and 3C~397, fall into the region of high ionization and high line luminosity, which is not covered by any of the models in this study.
The fact that these remnants cannot be explained despite the use of multiple explosion models indicates the necessity of considering alternative circumstellar environments beyond the uniform ISM assumed in the present simulations.
For SNR~0519–69.0, HST–Chandra–Spitzer comparisons reveal a velocity–brightness anti-correlation consistent with a dense, inhomogeneous CSM \citep{williams2022evidence}, and multi-epoch Chandra expansion measurements support the presence of a massive circumstellar shell \citep{guest2023rapid}.
For N103B, X-ray imaging and spectroscopy reveal a double-ring morphology consistent with an hourglass cavity, indicating interaction with a dense, asymmetric CSM \citep{yamaguchi2021discovery}.
For 3C~397, \textit{Chandra} imaging spectroscopy indicates evolution in a medium with a marked density gradient and a box-like morphology, pointing to a non-uniform environment \citep{safi2005chandra}. Broad-band Suzaku spectroscopy further suggests a near-$M_\text{Ch}$ Type Ia explosion in a high-density ambient medium \citep{martinez2020evidence}.
A most recent deep \textit{XMM-Newton} spatially resolved spectroscopic study compares the fitted metal abundances with a suite of thermonuclear and core-collapse nucleosynthesis models and shows that (a) no model fully reproduces the complete set of observed abundance patterns, and (b) the high Fe enrichment and spatial abundance variations suggest interaction with a dense progenitor environment \citep{2025arXiv251201176T}. 
Furthermore, the Fe-K line centroid was shown to span the thermonuclear plus CC energy region, further supporting the scenario of 3C~397 expanding into a highly non-uniform medium while underscoring the need for caution when using the Fe-K line centroid of average spectra for typing \citep{2025arXiv251201176T}.
We note that Fe-K$\alpha$ centroid and luminosity are shaped by both explosion-dependent properties (e.g., iron yield and the amount of shocked iron, as discussed below) and the ambient medium, including possible 3-D CSM structures. 
While environmental variations can, in principle, produce substantial shifts via changes in the ionization timescale and EM, reproducing the full observed spread especially in luminosity through the environment alone for a single explosion model would likely require exploring environmental parameters beyond a plausible range for the typical Type Ia environments.
Moreover, the explosion channel and environment may not be independent: different progenitor scenarios can plausibly lead to different degrees of CSM structuring, which can further broaden the range of Fe-K$\alpha$ properties.
In the uniform-ISM 1-D study of \citet{yamaguchi2014discriminating}, varying the ambient density for a single explosion model does not span the full observed region of the Fe-K$\alpha$, and multiple Type Ia explosion models are likely required.

The diversity in centroid energy depends on the explosion properties, while the diversity in line luminosity depends on both the ambient density and the fraction of IGE (see Table~\ref{tab:modelparameter}).
These features can be interpreted using the figures presented in Section~\ref{sec:plasma}. The high ionization state in the deflagration models is consistent with Figure~\ref{fig:net}, as iron is shocked at higher densities due to their slower expansion. Since the EM scales with the square of the density, the line luminosity can be relatively high even in the deflagration models, which have lower ejecta masses.

The evolution of line luminosity is also related to the amount of shocked mass. 
Notably, the \texttt{N100ddt} and \texttt{TwoExp} models show a substantial increase. 
As seen in Figure~\ref{fig:unshocked_mass_frac}, the unshocked mass fraction of Fe in these models is nearly unity at 100 years, but by 400 years, it decreases to approximately 0.35 in \texttt{N100ddt} and 0.6 in \texttt{TwoExp}, indicating significant shock processing of Fe. 
Moreover, in the \texttt{TwoExp} model, a large amount of iron becomes shocked between 400 and 700 years, consistent with the increase in line luminosity observed over the same period in Figure~\ref{fig:unshocked_mass_frac}. 
This is due to the impact of the shock wave from the secondary explosion, which propagates into and shocks the primary ejecta.

Thus, the characteristics of the Fe-K$\alpha$ line reflect the explosion properties, suggesting that comparisons between simulations and observed Fe-K$\alpha$ features can serve as a diagnostic to constrain explosion models.

\subsubsection{Volume integrated spectra}

\begin{figure*}[ht!]
\includegraphics[width=\textwidth]{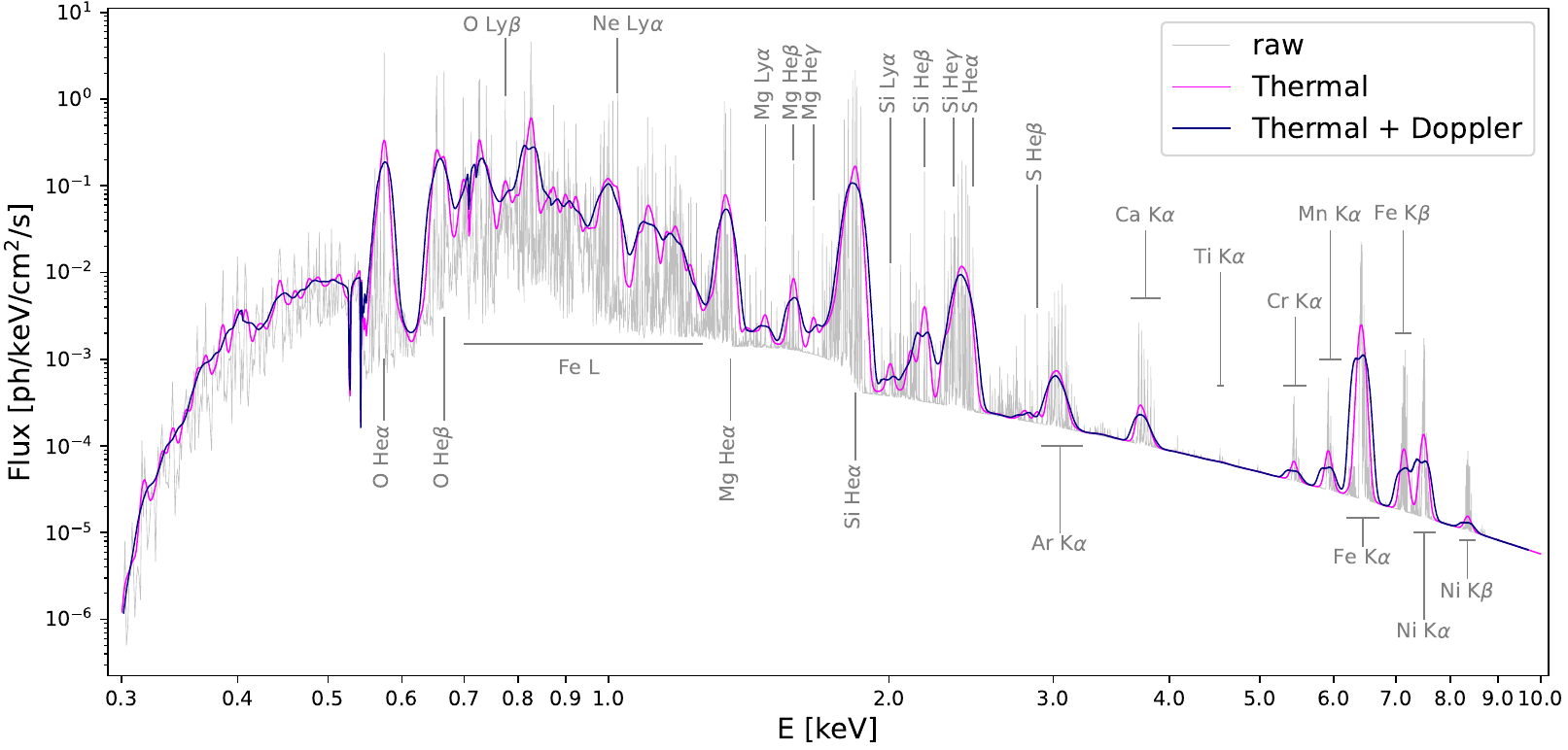}
\caption{Illustration of the synthesis of the volume-integrated broadband X-ray spectrum from our SNR models. Here we are showing an example using the \texttt{N100ddt} model at 400 years. The silver line shows the raw spectrum before applying the effects of thermal broadening and Doppler shift. The magenta line shows the spectrum with thermal broadening taken into account. The navy line shows the final spectrum with Doppler effect from bulk motion also accounted for in addition.
The absorption features around 5.5\,keV arise from spectral calculation settings, such as distance, column density and absorption model.
\label{fig:spec_N100ddt_raw}}
\end{figure*}

Figure~\ref{fig:spec_N100ddt_raw} is an illustration of the synthesis of the volume-integrated broadband X-ray spectrum from our SNR models. 
The silver line shows the raw spectrum before applying the effects of thermal broadening and Doppler shift.
This spectrum reveals the emission from each ionization state of each element as calculated by the NEI model.
The magenta line shows the spectrum with thermal broadening, and thus exhibits a symmetric broadening profile.
At 400 years, the ion temperature is high, so individual emissions from each ionization state are no longer distinguishable once thermal broadening is applied. 
The navy line shows the final spectrum with the Doppler effect.
For elements with asymmetric spatial distributions, the resulting line profiles exhibit asymmetric structures.
Even for elements with spherically symmetric distributions, the line width becomes broader than the spectrum with only thermal broadening due to the velocity distribution.
This comparison clarifies how each physical effect shapes the emergent spectrum, and highlights that accurate interpretation of X-ray spectra requires modeling both thermal broadening and Doppler shift in a fully 3-D framework.
We note that \citet{mandal2025snr} performed a 3-D hydrodynamic simulation based on 1-D initial conditions to investigate the evolution of turbulence, and their results supported a double detonation origin for SNR 0509–67.5. 
Given the intrinsic asymmetry of double detonation models (e.g., \texttt{OneExp} and \texttt{TwoExp}), the fully 3-D framework adopted in this study is expected to offer further insight through synthetic X-ray spectra, providing a more direct comparison with observations.

\begin{figure*}[ht!]
\includegraphics[width=\textwidth]{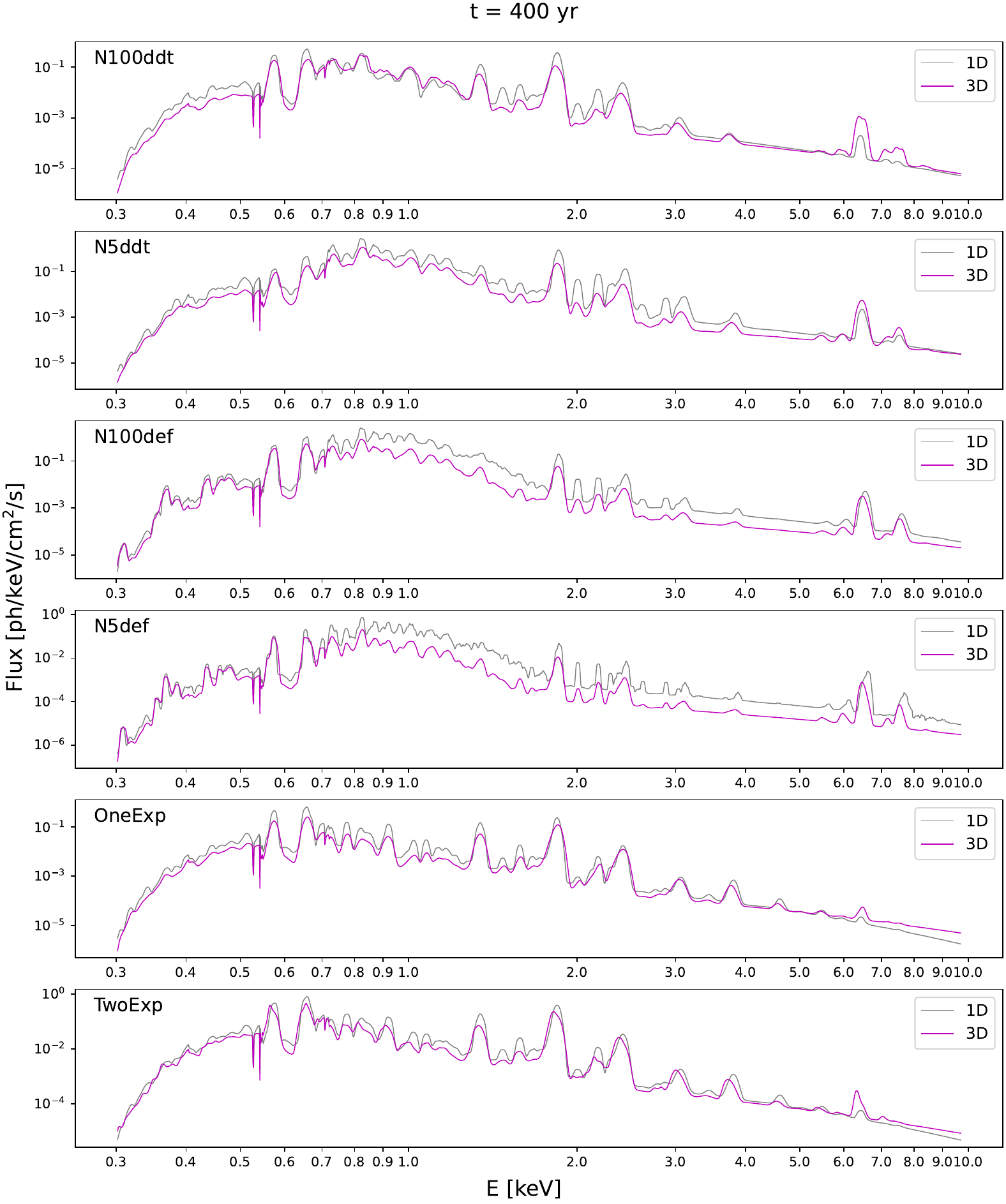}
\caption{A comparison between the volume-integrated broadband spectra of shocked ejecta at 400 years from the six 3-D SNR models and their averaged, spherically symmetric 1-D versions. The 1-D hydrodynamic simulations are performed by first deriving spherically symmetric initial conditions from averaging over the 3-D data, followed by a synthesis of their X-ray spectra using a method identical to the 3-D case. Both the 1-D and 3-D model spectra incorporate the effects of thermal broadening and Doppler shift. Axes are defined arbitrarily without physical directional meaning. We arbitrarily set the line-of-sight direction to Y for DDT and DEF models, and Z for DD models.
\label{fig:spec_1Dvs3D}}
\end{figure*}

Figure~\ref{fig:spec_1Dvs3D} shows broadband spectra of shocked ejecta at 400 years from the six 3-D SNR models and their averaged, spherically symmetric 1-D version.
The overall spectral features are broadly consistent between the 3-D and 1-D cases; similar emission lines are present, and the elemental abundances remain essentially unchanged after averaging, resulting in only minor differences.
Nevertheless, the neglect of spatial distributions in both position and velocity in the 1-D models introduces systematic differences that become apparent upon closer inspection.

The differences between the 1-D and 3-D spectra can be broadly divided into two aspects.
The first is the electron temperature, which is reflected in the continuum slope. 
The continuum is flatter in the 3-D spectra, indicating a higher electron temperature.
This is plausibly attributed to the clumpy structure of the 3-D SNR models, where higher-density regions tend to develop higher electron temperatures.
Since the emission measure is proportional to the square of the density, radiation from such high-temperature, hot clumps dominates the spectrum.
As a result, the electron temperature inferred from the observed X-ray spectrum is higher in the 3-D case.

The second aspect is the characteristic structure of the emission lines.
The differences lie in the shapes of the emission lines and their centroid energies.
In the 1-D models, the velocity distribution is spherically symmetric, so even when Doppler broadening is considered, the lines only widen without developing asymmetric structures.
In addition, slight differences in plasma evolution can cause variations in the ionization state, leading to small shifts in line positions.

A concrete example is provided by the Fe lines in the \texttt{N100ddt} model. 
The 3-D spectrum is noticeably brighter than its 1-D counterpart, reflecting the full 3-D spatial distribution of Fe. 
While Fe is initially concentrated in the central region, the 3-D models reveal that some Fe clumps extend to larger radii and are shocked earlier. 
In the 1-D case, spherical averaging cancels out such protruding structures, thereby underestimating the contribution from shocked Fe.
These comparisons show that, even when global spectra look similar, only 3-D models can fully capture the underlying ejecta structure and dynamics.

\begin{figure*}[ht!]
\includegraphics[width=0.99\textwidth]{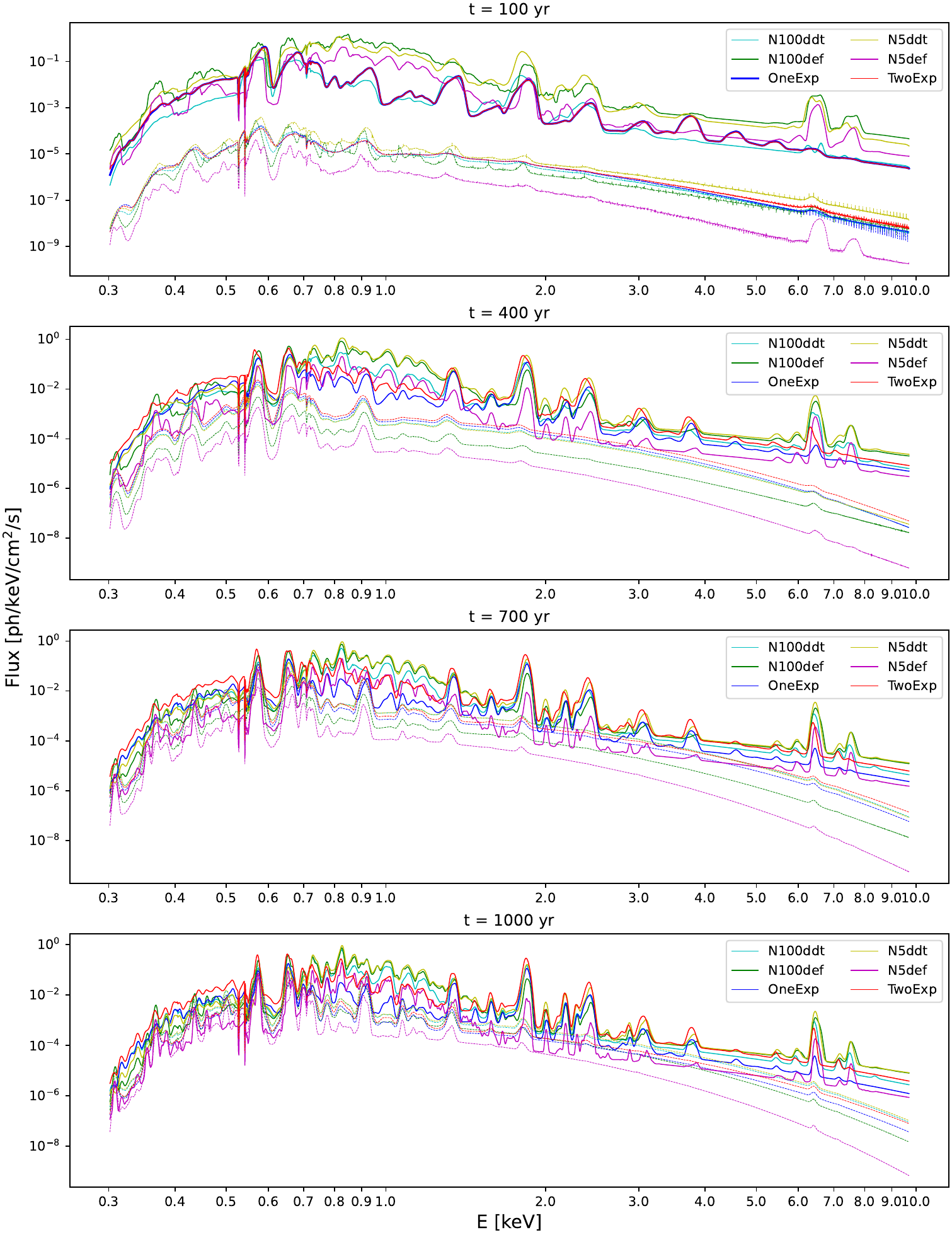}
\caption{A comparison of the volume-integrated X-ray spectra from the six explosion models at four SNR ages. The solid lines show the spectrum from the shocked ejecta. The dashed lines show the spectrum from the shocked ISM. The line-of-sight directions are the same as those in Figure~\ref{fig:spec_1Dvs3D}. In some ISM spectra, emission lines are seen around 6--8 keV. 
These features are artifacts caused by contamination from the ejecta. 
The flux of the ISM component is calculated as
$F_{\mathrm{ISM}} = F_{\mathrm{tot}} \times (1 - f_{\mathrm{ej}})$,
where $f_{\mathrm{ej}}$ denotes the ejecta fraction in each tracer particle.
Because both components are derived from the same tracer particles, minor contamination between them is unavoidable.
\label{fig:all_evo}}
\end{figure*}

Figure~\ref{fig:all_evo} shows the spectra of the six models at four different ages. The solid lines represent the spectra from the shocked ejecta, while the dashed lines correspond to those from the shocked ISM. From top to bottom, SNRs evolve with time.

The spectral evolution of the SNRs is characterized by systematic changes in line widths, asymmetries, and relative luminosities of ejecta and ISM components. 
At first, ejecta dominate the X-ray emission by factors of $\sim$100--1000, producing broad and asymmetric emission lines due to their high velocities, consistent with the error bars shown in Figure~\ref{fig:Yamaguchi}. 
By $\sim$1000 yr, the ejecta have decelerated sufficiently so that Doppler shifts become less effective and the emission lines appear narrower and more symmetric. 
Meanwhile, the ISM contribution increases steadily. 
In the \texttt{N100ddt} model, for example, the remnant radius grows from $\sim$1.5 pc at 100 years to $\sim$7 pc at 1000 years, increasing swept-up mass by about two orders of magnitude. 
As a result, the ISM component becomes comparable to ejecta emission at late times, consistent with the shift of EM-weighted peaks from ejecta (circle) to ISM (cross) in Figure~\ref{fig:net}.

Spectral differences between models are reflected in luminosity, line presence, and line morphology. 
The overall luminosity scales with the ejecta mass, since a larger ejecta mass yields a larger shocked mass.
The ejecta masses for each model are listed in Table~\ref{tab:modelparameter}.
An exception occurs in the \texttt{N100ddt} model, whose ejecta mass is $1.4\,\mathrm{M_{\odot}}$ but the luminosity at 100 years is relatively low, a feature also seen in Figure~\ref{fig:Yamaguchi} and attributable to its radial mass distribution and shock evolution. 
The emission line patterns also follow the initial abundance structures. 
In Figure~\ref{fig:abund_pattern_diff}, the shocked masses normalized to \texttt{N100ddt} correspond directly to the spectral features: DD models exhibit weaker IGE lines but stronger IME lines, reflecting their enhanced IME abundances and lower IGE yields due to the smaller progenitor mass.

In addition to these global features, certain lines provide distinctive diagnostics. 
The Ti abundances in DD models lead to faint emission around 4.5 keV, which is indeed visible in their spectra. 
Although weak, these lines could be detected with high-resolution spectroscopy from missions such as XRISM, providing valuable constraints on explosion models. 
Finally, as also discussed in Section~\ref{sec:region}, line profiles vary with the degree of asymmetry; analyzing Doppler shifts offers a means to probe the spatial distributions of individual elements.

\subsubsection{Region specific spectra and dependence on line-of-sight} \label{sec:region}

\begin{figure*}[ht!]
\includegraphics[width=0.98\textwidth]{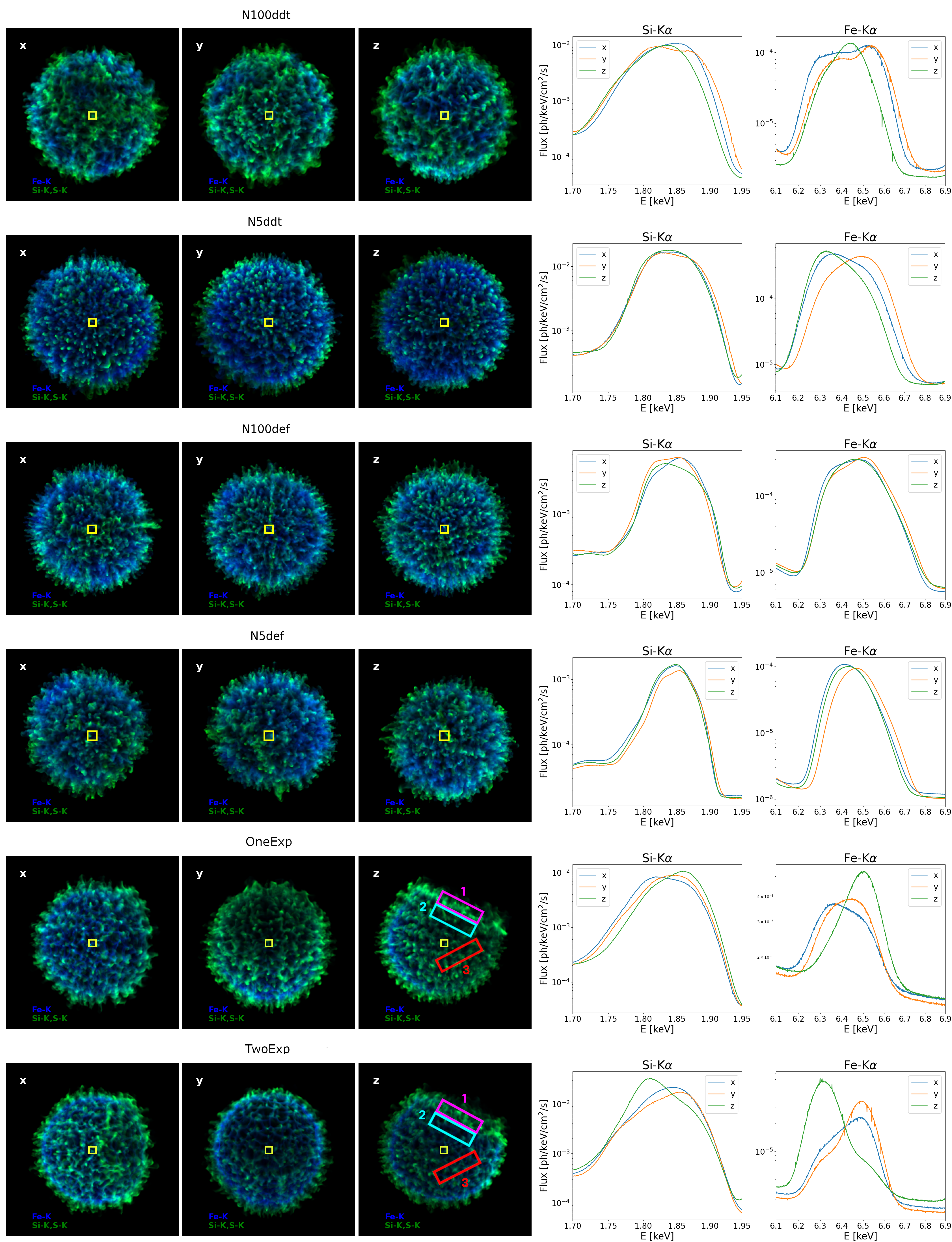}
\caption{
A comparison of line-of-sight specific emission line profiles of Si-K$\alpha$ and Fe-K$\alpha$ at 400 years. The spectra are extracted from a $0.5' \times 0.5'$ region (see  yellow square, approximately the pixel size of XRISM Resolve) at the center of each SNR for three line-of-sight directions. All the SNRs are assumed to be at a 2.5 kpc distance here. A stronger dependence of the profiles on the viewing angle can be seen from the Fe line than the Si line for some explosion models, showing the elemental dependence of the ejecta asymmetry. 
The images follow \citet{ferrand2021supernova, ferrand2025role}, and depict emissivity maps constructed in the energy band of each line.
\label{fig:all_center}}
\end{figure*}

Figure~\ref{fig:all_center} shows a comparison of line-of-sight specific emission line profiles of Si-K$\alpha$ and Fe-K$\alpha$ at 400 years. 
The spectra are extracted from a $0.5' \times 0.5'$ region (see yellow square, approximately the pixel size of XRISM Resolve) at the center of each SNR for three line-of-sight directions.
Since the central region contains a large fraction of line-of-sight velocity components, the effects of red-shifted and blue-shifted emission can be clearly observed. 
In addition, restricting the region avoids the averaging of asymmetric features and highlights asymmetries of spectra.
Moreover, by focusing on the central region, emission from all radial layers can be observed. This makes it particularly useful for gaining an overview of the explosion model.

As a general trend, spherically symmetric models exhibit broad emission line structures that resemble a hill-like shape or double-peaked profiles. 
In contrast, models with strong asymmetry show line profiles that are shifted to either side. 
At 400 years, the high ion temperatures result in strong thermal broadening, which smears out fine structures. 
Therefore, these spectral features are considered to primarily reflect the spatial distribution.

In the delayed-detonation models (top two panels), the effect of Doppler shift is clearly seen in the Fe-K$\alpha$ emission lines, whereas it is less apparent in the Si-K$\alpha$ lines. This difference can be attributed to the fact that the velocity required for the Doppler shift to dominate differs between the two lines, as discussed in the plasma diagnostics presented in Figure~\ref{fig:T_v_Si_400} and Figure~\ref{fig:T_v_Fe_400} of Section~\ref{sec:plasma}. 

The two delayed-detonation models also exhibit differences in the Fe distribution, which can be inferred from the spectrum profiles.
In the \texttt{N100ddt} model, spectra along the X and Y directions show a double-peaked Fe-K$\alpha$ line structure, indicating a roughly spherically symmetric Fe distribution, where the red-shifted and blue-shifted components in the central region are comparable.
In contrast, the \texttt{N5ddt} model exhibits red-shifted profiles along the X and Z directions, and a blue-shifted profile along the Y direction. This suggests that Fe is asymmetrically distributed along specific directions, consistent with the asymmetry in the explosion itself.

The deflagration models exhibit smaller variations among different line-of-sights compared to the other models. 
This can be attributed to the relatively low velocities of Si and Fe.
As shown in Table~\ref{tab:modelparameter}, the deflagration models have comparatively lower kinetic energies, and Figure~\ref{fig:T_v_Si_400} and Figure~\ref{fig:T_v_Fe_400} demonstrate that their velocity distributions are narrower than those of the other models, accounting for the lower expansion velocities.
Nevertheless, a comparison between \texttt{N100def} and \texttt{N5def} indicates that the latter shows larger variations across line-of-sights. 
This can be explained by the smaller number of ignition points, which leads to a more asymmetric explosion.
As shown in Figure~\ref{fig:T_v_Si_400} and Figure~\ref{fig:T_v_Fe_400}, the ejecta in \texttt{N5def} are confined to low-velocity regions, resulting in even narrower emission line profiles.
Focusing on \texttt{N5def} of the Figure~\ref{fig:T_v_Fe_400}, one can see that the regions with high EM are concentrated on one side, which accounts for the variations between different lines of sight.

The DD models show the most pronounced asymmetries. 
These features can be attributed to the presence of the companion WD, which produces a shadow-like structure behind it, enhancing the anisotropy.
Following \citet{ferrand2022double,ferrand2025role}, we use the term ``shadow'' to refer to the geometrical, conical low-density region (often described as an ejecta ``hole'') formed behind the secondary WD as it blocks the expanding ejecta, and we do not imply any optical shadowing or attenuation of the observed X-ray emission.
The other models considered here do not include an explicit companion; if a companion were present, similar shadow-like signatures would likely appear.
The comparison between \texttt{OneExp} and \texttt{TwoExp} is particularly striking: While both show strong asymmetries, the directions of Doppler shifts differ. 
In the \texttt{TwoExp} model, the explosion of the secondary alters the orientation of the asymmetry compared to \texttt{OneExp}, further emphasizing the effect of the binary interaction.

\begin{figure*}[ht!]
\includegraphics[width=\textwidth]{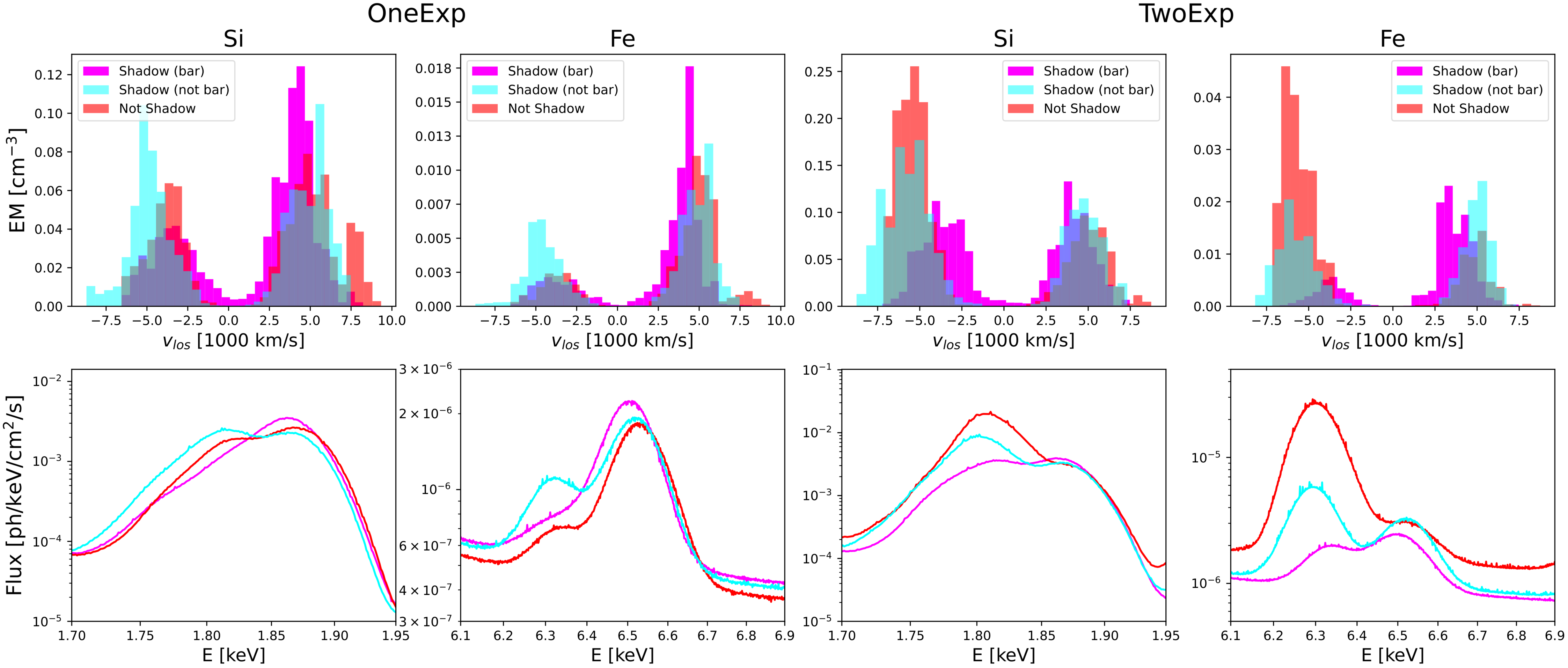}
\caption{Top panel: The distribution of EM of extracted regions (see Figure~\ref{fig:all_center}) in the velocity space. (1: Shadow (bar), 2: Shadow (not bar), 3: Not Shadow) Here, $v_\mathrm{los}$ is the projected velocity of Si and Fe along Z-axis. Bottom panel: Same as right panels of Figure~\ref{fig:all_center}, but for regions other than the center, which correspond to those shown in the top panels
\label{fig:bar_shadow}}
\end{figure*}

These results demonstrate that spectra extracted from central regions provide powerful diagnostics of explosion asymmetry. 
In Figure~\ref{fig:bar_shadow}, we explore spectra from specific off-center regions in the \texttt{OneExp} and \texttt{TwoExp} models to assess how local variations further constrain the explosion mechanisms.

Figure~\ref{fig:bar_shadow} shows the distribution of the emission measure (EM) of the extracted regions in the velocity space and the Si-K$\alpha$ and Fe-K$\alpha$ emission lines from these regions.
The most prominent difference between the \texttt{OneExp} and \texttt{TwoExp} models appears in the velocity distributions and emission lines of the not-shadow regions.
Inspection of the spectra and velocity distributions in this region reveals that, while the \texttt{OneExp} model exhibits only a small fraction of redshifted components, the \texttt{TwoExp} model shows a significantly larger redshifted contribution.
This can be attributed to the explosion of the secondary WD.
Thus, analyzing spectra and velocity distributions on a region-by-region basis allows us to qualitatively identify differences between the models.
Such spatially resolved spectral analysis is precisely the method enabled by the XRISM mission, demonstrating that our simulations provide a valuable tool for constraining explosion models.

\section{Summary} \label{sec:summary}

Type Ia supernova remnants show significant diversity in their X-ray emission, especially Fe-K$\alpha$ emission.
To investigate whether this diversity can be explained by differences in explosion mechanisms, we performed fast 3-D hydrodynamic simulations by using six explosion models, followed self-consistently to 1000 years including NEI. 
The ambient medium was set to $n_\text{ISM}=0.3$ cm$^{-3}$, consistent with that inferred for Tycho’s SNR. 
Subsequently, we calculated X-ray spectra with an energy resolution of $\sim$1 eV, enabling direct comparison with XRISM observations.

The main results are as follows:
\begin{itemize}
\setlength{\itemsep}{-5pt}
 \item The hydrodynamical simulations reveal diversity in ionization evolution, which correlates with explosion parameters such as ejecta mass and explosion energy.
 
 \item The distributions of velocity and temperature have differences among the models: more asymmetric explosions exhibit stronger biases in the velocity space. Even within the same model, different elements show different distributions, highlighting the importance of 3-D effects.
 
 \item The shocked mass fractions and absolute masses of individual elements evolve differently in the models, and their evolutions did not simply scale with time but reflect element-dependent 3-D distributions. This feature is crucial when comparing to observations of young SNRs.
 
 \item The properties of the Fe-K$\alpha$ line reproduce much of the observed diversity of Type Ia SNRs, with the models roughly clustering into three regions corresponding to the explosion mechanisms.

 \item However, the most highly ionized and luminous Fe-K$\alpha$ sources (0519--69.0, N103B, and 3C~397) cannot be reproduced under our uniform ISM assumption of $n_{\mathrm{ISM}} = 0.3~\mathrm{cm}^{-3}$.
 This indicates that explosion model diversity in a single density ISM is not sufficient to explain the full observed diversity of Type Ia SNRs, and that variations in the ambient ISM density and asymmetric 3-D CSM environments must also be taken into account.
 
 \item Global spectra show significant model-to-model diversity. Emission lines of Si, S, Ca, Ti, Cr, Mn, Fe and Ni appear in the synthetic spectra, but some lines are present only in specific models, providing potential diagnostics for explosion mechanisms. Time evolution further reveals that the contributions of ejecta and ISM become comparable by $\sim$1000 years.

 \item Line profiles extracted from the central region, on a scale corresponding to XRISM’s pixel size, exhibit variations with line of sight. Asymmetric models in particular show skewed or shifted line structures, indicating that such profiles can be used to diagnose explosion asymmetry and direction of line-of-sight.

 \item In double detonation models, line profiles from specific regions also differ. In some cases, the redshift/blueshift direction is reversed between models for the same region, and even within a single model, the redshift/blueshift ratios vary across different regions.
 
\end{itemize}

This study presents, for the first time, a fully self-consistent calculation of the evolution from 3-D Type Ia supernova explosion models to 3-D supernova remnants, together with the associated X-ray spectra. 
This approach enabled us to directly trace differences in abundances and 3-D structures among explosion models. 
The resulting spectral variations are in principle testable with X-ray spectroscopy, and qualitative constraints on explosion models can be obtained by comparing these simulations with observations.
Tycho (SN 1572) is a representative Type Ia SNR whose nearly spherical morphology and kinematics are consistent with evolution in a roughly uniform ISM. 
High-resolution spectroscopy with XRISM now provides precise line profiles and abundances for Tycho; confronting these data with our synthetic spectra offers a promising route to discriminate progenitor models. 
A dedicated observation–model comparison is left for future work.

To isolate the effects of explosion models, we adopted a uniform ISM environment in this work. 
However, certain regions of Figure~\ref{fig:Yamaguchi}, such as the high ionization and high luminosity region seen in 0519–69.0, N103B, and 3C~397, could not be reproduced.
These cases are likely associated with SNRs that interact with a dense CSM.
An interesting example is W49B (not shown in Figure~\ref{fig:Yamaguchi}, lying in the CC region of \citealt{yamaguchi2014discriminating}), which has been considered to originate from a CC SN due to its high ionization state and bipolar morphology.
However, spatially resolved X-ray spectroscopy with Chandra revealed element abundances and spatial distributions consistent with asymmetric Type Ia explosion models \citep{zhou2018asymmetric}. 
This interpretation was further supported by Suzaku spectra showing a low Ti/Fe mass ratio \citep{sato2025missing}, and most recently by XRISM observations, which confirmed Type Ia-like abundance patterns and detected Doppler shifts indicative of a bar-like structure \citep{xrism2025kinematic}.
Most recently, \citet{2025arXiv251201176T} compared the inferred metal abundance ratios (using a spatially resolved \textit{XMM-Newton} study) to a full suite of core-collapse and thermonuclear nucleosynthesis models, and found that the observed Fe/Si and Ca/Si ratios are best matched by certain thermonuclear models; however, no model fully reproduces the complete set of observed abundance patterns. 
The Fe K$\alpha$ line centroid energies yielded a spread, with W49B falling within the CC region (see also \citet{2020ApJ...904..175S}), again presenting tension between the nucleosynthesis model typing and the Fe~K$\alpha$ line centroid methodology as demonstrated for 3C~397 \cite{2025arXiv251201176T}.
Explaining such SNRs within the framework of Type Ia explosion models requires incorporating highly asymmetric 3-D CSM environments.
Future studies should therefore extend this framework to 3-D CSMs and incorporate a broader range of explosion scenarios.
In addition, to facilitate direct comparison with XRISM observations, we will include observation-oriented simulations that account for instrumental properties, such as exposure time.
Looking ahead, next-generation missions such as \textit{AXIS} \citep{2023SPIE12678E..1ER,2025arXiv251100253K,2023arXiv231107673S} and \textit{NewAthena} \citep{cruise2025newathena} will offer high-spatial-resolution imaging and high-resolution X-ray spectroscopy. 
These capabilities will provide detailed spatial and spectral data of SNRs, requiring theoretical models that incorporate 3-D structures for a meaningful comparison.
The combination of such detailed observations and advanced 3-D models will further elucidate the evolution from SN phase to SNR phase.

\begin{acknowledgments}

H.L. acknowledges support from JSPS grant No. JP19K03913. S.S.H. acknowledges support from the Natural Sciences and Engineering Research Council of Canada (NSERC) through the Canada Research Chairs and Discovery Grants Programs. Supercomputing support in Canada is provided by the Digital Research Alliance of Canada. S.N. is supported by JSPS Grant-in-Aid Scientific Research (KAKENHI) (A), Grant Number JP25H00675, and (B), Grant Number JP23K25874, and JST ASPIRE Program “RIKEN-Berkeley mathematical quantum science initiative”. The work of F.K.R. is supported by the Klaus Tschira Foundation, the Deutsche Forschungsgemeinschaft (DFG, German Research Foundation) – RO 3676/7-1, project number 537700965, and by the European Union (ERC, ExCEED, project number 101096243). Views and opinions expressed are, however, those of the authors only and do not necessarily reflect those of the European Union or the European Research Council Executive Agency. Neither the European Union nor the granting authority can be held responsible for them. A.D. acknowledges support from the Centre National d’Etudes Spatiales (CNES). D.J.P. acknowledges support from the Chandra X-ray Center, which is operated by the Smithsonian Institution under NASA contract NAS8-03060.

\end{acknowledgments}

\appendix

\section{On the projected velocity of iron}
\label{sec:append_dist_vlos}

\renewcommand{\thefigure}{A\arabic{figure}}
\setcounter{figure}{0}

\begin{figure*}[ht!]
\includegraphics[width=\textwidth]{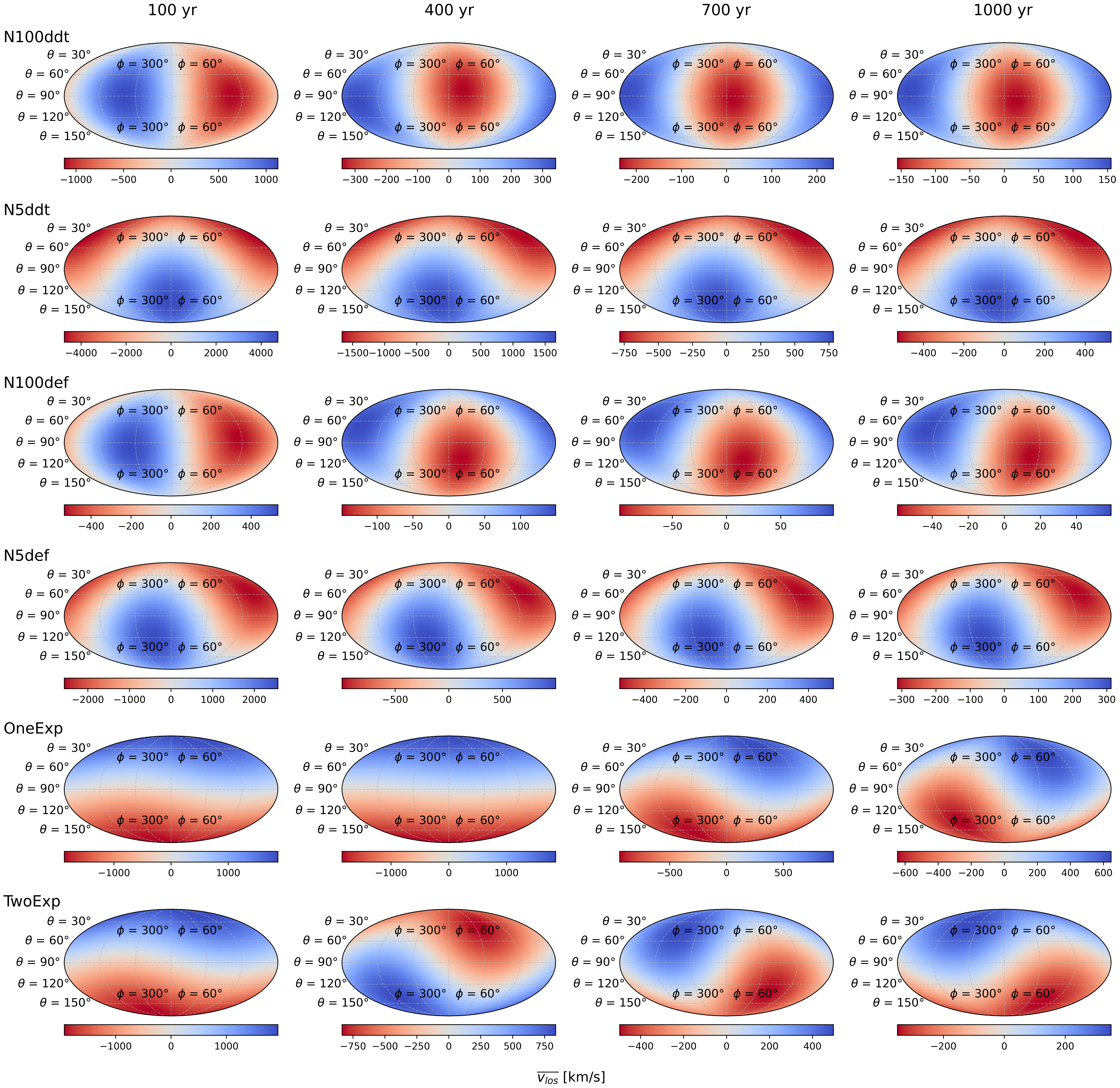}
\caption{Distribution of the EM-weighted projected velocity of Fe in the shocked ejecta as a function of viewing angle ($\theta$, $\phi$), for the six explosion models at four different ages. Abrupt transitions of the distribution between time epochs showcase drastic changes in the dynamics of the shocked Fe ejecta, which is especially obvious for the \texttt{TwoExp} model between the 100 and 700 year time stamps when the explosion of the secondary WD begins to affect the X-ray emitting ejecta.  
Positive velocities, weighted by the Fe emission measure, indicate that the Fe-emitting plasma is dominated by blueshifted components (shown in blue), while negative velocities indicate redshifted components (shown in red).
\label{fig:dist_vlos}}
\end{figure*}

Figure~\ref{fig:dist_vlos} show the distribution of the EM-weighted projected velocity of Fe in the shocked ejecta as a function of viewing angle $(\theta, \phi)$, for the six explosion models at four different ages.

We calculated average velocity as function of $\theta$ and $\phi$ by using following equation:
\begin{equation}
 \overline{v}_{(\theta, \phi)}=\frac{\sum_i \bm{n}_{(\theta, \phi)} \cdot \bm{v}_i \times EM_{Fe, i}}{\sum_i EM_{Fe, i}}
\label{eq:vlos}
\end{equation}
where $\bar{v}_{(\theta, \phi)}$ is average velocity along viewing angle $(\theta, \phi)$, $\bm{n}_{(\theta, \phi)}$ is unit vector along viewing angle $(\theta, \phi)$, $i$ is number of shocked tracer particle, $\bm{v}_i$ is velocity of particle $i$ and $EM_{Fe, i}$ is emission measure of Fe of particle $i$.
Although the maps may appear to show different structures for each model, this is mainly due to the definitions of $\phi$ and $\theta$.
The key features to focus on are the changes in global trends with time and the maximum value of the color bar.
The latter corresponds to the average velocity along the direction with the largest $v_{\mathrm{los}}$, which tends to be higher for models with higher explosion energies.
As time progresses, Fe closer to the center becomes shocked, leading to a decrease in velocity.

Such global changes are most clearly seen in the \texttt{N100ddt}, \texttt{N100def} and \texttt{TwoExp} models, and to a lesser extent in \texttt{OneExp}.
For \texttt{N100ddt} and \texttt{N100def}, as shown in Figure~\ref{fig:unshocked_mass_frac}, most of the ejecta are shocked between 100 and 400 years, shifting the dominant region toward the center.
In the case of \texttt{TwoExp}, the velocity structure is modified by the secondary explosion, while in \texttt{OneExp} the asymmetry arises from the influence of the companion WD.

This type of analysis is important for investigating the 3-D distribution of elements and for evaluating their impact on the resulting spectra.

\bibliography{references}{}
\bibliographystyle{aasjournal}

\end{document}